\documentclass{aa}  

\newcommand{\Teff}{T_{\rm eff}}

\newcommand{\FeH}{[{\rm Fe/H}]}
\newcommand{\OFe}{[{\rm O/Fe}]}
\newcommand{\MgFe}{[{\rm Mg/Fe}]}
\newcommand{\SiFe}{[{\rm Si/Fe}]}
\newcommand{\CaFe}{[{\rm Ca/Fe}]}
\newcommand{\AFe}{[{\rm \alpha/Fe}]}
\newcommand{\AM}{[{\rm \alpha/M}]}

\usepackage{natbib}
\bibpunct{(}{)}{;}{a}{}{,} 
\usepackage{graphicx}
\usepackage{txfonts}
\begin{document}

   \title{Radial structure and formation of the Milky Way disc}

   \subtitle{}

   \author{D. Katz
          \inst{1}
          \and
          A. G\'omez\inst{1}
	  \and
	  M. Haywood\inst{1}
          \and
          O. Snaith\inst{1}
          \and
	  P. Di Matteo\inst{1}
          }

   \institute{GEPI, Observatoire de Paris, Universit\'e PSL, CNRS, 5 Place Jules Janssen, 92190 Meudon, France\\
              \email{david.katz@obspm.fr}}

   \date{Received ; accepted }

 
  \abstract
   {The formation of the Galactic disc is an enthusiastically debated issue. Numerous studies and models seek to identify the dominant physical process(es) that shaped its observed properties: e.g. satellite accretion, star burst, quenching, gas infall, stellar radial migration.}
   {Taking advantage of the improved coverage of the inner Milky Way provided by the SDSS DR16 APOGEE catalogue and of the ages published in the APOGEE-AstroNN Value Added Catalogue (VAC), we examine the radial evolution of the chemical and age properties of the Galactic stellar disc, with the aim to better constrain its formation.}
   {Using a sample of 199,307 giant stars with precise APOGEE abundances and APOGEE-astroNN ages, selected in a $\pm2$~kpc layer around the galactic plane, we assess the dependency with guiding radius of: (i) the median metallicity, (ii) the ridge lines of the $\FeH$-$\MgFe$ and age-$\MgFe$ distributions and (iii) the Age Distribution Function (ADF).}
   {The giant star sample allows us to probe the radial behaviour of the Galactic disc from $R_g =$~0 to 14-16~kpc. The thick disc $\FeH$-$\MgFe$ ridge lines follow closely grouped parallel paths, supporting the idea that the  thick disc did form from a well-mixed medium. However, the ridge lines present a small drift in $\MgFe$, which decreases with increasing guiding radius. At sub-solar metallicity, the intermediate and outer thin disc $\FeH$-$\MgFe$ ridge lines follow parallel sequences shifted to lower metallicity as the guiding radius increases. We interpret this pattern, as the signature of a dilution of the inter-stellar medium from $R_g \sim 6$~kpc to the outskirt of the disc, which occured before the onset of the thin disc formation. The APOGEE-AstroNN VAC provides stellar ages for statistically significant samples of thin disc stars from the Galactic centre up to $R_g \sim 14$~kpc. An important result provided by this dataset, is that the thin disc presents evidence of an inside-out formation up to $R_g \sim 10-12$~kpc. Moreover, about $\sim 7$~Gyr ago, the $\MgFe$ ratio in the outer thin disc ($R_g > 10$~kpc) was higher by about $\sim 0.03-0.05$~dex than in the more internal regions of the thin disc. This could be the fossil record of a pollution of the outer disc gas reservoir, by the thick disc during its starburst phase.}
   {}

   \keywords{The Galaxy -- Galaxy: disk -- Galaxy: evolution -- Galaxy: abundances}

   \maketitle


\section{Introduction}
The Milky Way stellar disc is a complex structure. Its formation and its  evolution together with the related physical and dynamical processes are still strongly debated. Recently, several surveys have provided very large statistical samples of stars
with detailed chemical abundances: e.g. RAVE \citep{Steinmetz2020_1, Steinmetz2020_2}, Gaia-ESO \citep{Gilmore2012}, APOGEE \citep{Majewski2017, Ahumada2020, Jonsson2020}, LAMOST \citep{Cui2012, Zhao2012} or GALAH  \citep{DeSilva2015, Buder2018, Buder2020}. Combined with the very precise Gaia astrometry \citep{GaiaCollaboration2018, GaiaCollaboration2020, Lindegren2018, Lindegren2020a, Lindegren2020b} they allow to examine the formation of the Galactic disc with an unprecedented level of details. However, despite this amount of data, the origin and the link between the two traditionally assumed components of the Milky Way disc, i.e. the thick and the thin discs, are not well established. 

Spectroscopic studies of Solar neighbourhood stars showed that thick disc stars are generally older and have higher ratios of $\alpha$-elements than thin disc stars \citep[e.g.][]{Fuhrmann1998, Fuhrmann2004, Fuhrmann2008,  Fuhrmann2011, Prochaska2000, Bensby2005, Bensby2007, Haywood2013}.The star distribution in the ($\FeH$,$\AFe$) plane displays two sequences, a high-$\AFe$ sequence associated with the thick disc and a low-$\AFe$ sequence associated with the thin disc \citep{Fuhrmann1998, Haywood2013}. Moreover, the  velocity  dispersions are larger in the thick disc than in the thin disc and, as a whole, the thick disc rotates more slowly than the thin disc \citep[e.g.][]{Soubiran2003, Kordopatis2013, Robin2017}.  
Although the two discs would appear to belong to different stellar populations, some studies suggested that these populations are composed of multiple sub-populations that smoothly span the observed range of properties \citep{Norris1987, Bovy2012, Bovy2016, Mackereth2017}.

Beyond the Solar vicinity, the thin and thick disc stars still appear as two distinct groups in the ($\FeH$, $\AFe$) plane \citep[e.g.][]{Nidever2014, Hayden2015, Queiroz2020}, but their pattern evolves with the Galactic location ($R$, $|Z|$)\footnote{where $R$ and $Z$ are two of the Galactocentric cylindrical coordinates, respectively the projected distance to the Galactic centre in the Galactic plane and the distance perpendicular to the Galactic plane}. In the rest of the introduction, we will use the term $\alpha$-dichotomy to refer to the presence of two distinct, high- and low-$\alpha$ sequences and the term $\alpha$-bimodality to refer explicitly to the existence of two density peaks along a single sequence. The high-$\AFe$ sequence is observed up to $R \sim$10-12 kpc and gradually vanishes at larger radii, confirming that the thick disc has a shorter scale length than the thin disc \citep{Bensby2011, Bovy2012, Cheng2012}. The low-$\AFe$ sequence is present at all radii but its morphology changes with the location in the Galaxy. In the inner disc (R $\leq$ 6 kpc), \citet{Hayden2015, Bovy2019} and \citet{Lian2020c, Lian2020d} found the thick and thin disc stars to belong to a single sequence (i.e. $\alpha$-bimodality), while \citet{RojasArriagada2019} and \citet{Queiroz2020} observed two distinct sequences (i.e. $\alpha$-dichotomy). Moreover, the observed metallicity gradient in the thin disc is flat in the inner disc in contrast with the gradient observed in the external regions (R$>$ 6kpc) suggesting a different chemical evolution of the two regions \citep{Haywood2019}.

Different approaches exist in the literature to explain the formation paths of the $\alpha$-dichotomy and $\alpha$-bimodality seen in the  Galactic disc, and the formation of the thick disc. Among the chemical evolution models, one of the first approaches seeking to explain first the metallicity-$\alpha$-elements distribution in the solar vicinity \citep{Chiappini1997, Chiappini2001} and then the $\alpha$-dichotomy \citep[e.g.][]{Spitoni2019, Spitoni2020} is the two-infall model. The thick disc is formed at early times as the result of a collapse of primordial gas followed by a quiescent period where the star formation is quenched. After that, a new episode of fresh gas accretion takes place, the star formation resumes and the thin disc is formed in an inside-out fashion. A recent approach which also consists in two phases of gas accretion was developed by \cite{Lian2020b, Lian2020a, Lian2020c, Lian2020d} to reproduce the $\alpha$-dichotomy and $\alpha$-bimodality observed at different Galactic locations. The two gas accretion phases are associated with star bursts separated by a prolonged period of low-level star formation. In \citet{Haywood2018, Haywood2019} the thick disc forms from a turbulent gas-rich environment which is followed by a low level of star formation. The inner thin disc is formed from the gas left by the thick disc formation and there is no gas accretion. \cite{Snaith2015} showed that the inner and outer discs do not follow the same chemical evolution. The $\alpha$-bimodality observed in the inner disc ($R \leq$6 kpc) is an effect of the inner disk evolution and comes from the quenching phase that occurred between the formation of the thick disc and the inner thin disc \citep{Haywood2016}. In the external disc the low-$\alpha$ sequence is the result from the gas left over from the thick disk formation (which gave rise to the high-$\alpha$ sequence) combined with a radially dependent dilution \citep{Haywood2019}. In their scenario, the sequence itself cannot be reproduced by chemical evolution at a single radius, but is the result of a radially dependent evolution.

Recently, \citet{Khoperskov2021} presented a set of chemo-dynamical Milky Way-type galaxy formation simulations. In these simulations, the high-$\AFe$ sequence is formed early from a burst of star formation in a turbulent, compact gaseous disc which forms a thick disc. The low-$\AFe$ sequence, in turn, is the result of a quiescent star formation supported by the slow accretion of enriched gas onto a radially extended thin disc. Stellar feedback-driven outflows during the formation of the thick disc are responsible for the enrichment of the surrounding gaseous halo, which subsequently feeds the disc on a longer time-scale. \citet{Khoperskov2021} confirm the scheme presented in \citet{Haywood2019} by showing that the formation of the low-$\alpha$ sequence is a radially dependent process, which explains the $\alpha$-dichotomy. These simulations do not include mergers, which therefore did not lead here to the assembly of the $\alpha$-dichotomy.

\citet{SchonrichBinney2009a, SchonrichBinney2009b} developed a chemical evolution model which reproduced the $\alpha$-dichotomy in the ($\FeH$-$\AFe$) plane observed in the Solar neighbourhood assuming a continuous star formation.The model includes, in particular, radial flow of gas and radial migration of stars as the main mechanisms allowing to carry kinematically hot stars from the inner disc to the Solar vicinity and to create the thick disc. Recently, \citet{Sharma2020} confirmed the results of \citet{SchonrichBinney2009a, SchonrichBinney2009b} and extended them to different locations across the Galaxy.

Simulations of Milky Way mass galaxies with enough resolution have allowed to study the formation scenario of distinct discs self-consistently in a fully cosmological context. Such simulations predict that thin stellar discs form in an inside-out, upside-down fashion \citep[e.g.][]{Brook2012, Bird2013}, but the formation mechanism of the $\alpha$-dichotomy in the ($\FeH$-$\AFe$) plane is still under debate. Some authors \citep[e.g.][]{Brook2012, Grand2018, Noguchi2018, Buck2020, Agertz2020} have invoked gas-rich mergers as the main origin, but the formation channels of chemically distinct discs differ. For example \citet{Grand2018} found two main paths to produce the $\alpha$-dichotomy in different regions of the galaxies: an early centralised starburst mechanism that is relevant for the inner disc; and a shrinking disc mechanism that is relevant for the outer disc. For the centralised starburst pathway, an early and intense high-$\AFe$ star formation phase induced by gas-rich mergers is followed by a more quiescent low-$\AFe$ star formation. For the shrinking disc pathway, an early phase of high-$\AFe$ star formation is followed by a shrinking of the gas disc due to a temporarily lowered gas accretion rate, after which disc growth resumes. Otherwise, \citet{Clarke2019} claimed that the dichotomy arises from rapid star formation in high redshift clumps and the two $\AFe$-sequences form simultaneously early on in the evolution of the Milky Way, resulting in overlapping ages.

The time evolution of radial chemical abundance gradients \citep{Anders2017} and of the $\AFe$ and $\FeH$ abundances \citep[e.g.][]{Haywood2013, Xiang2017, SilvaAguirre2018, Wu2019, Miglio2020, Ciuca2020} are crucial to explore the formation history of the Galaxy. Age is not a directly observable quantity. Asteroseismology and high-spectroscopic data allowed to measure precise masses of tens of thousands of red-giants and then to derive stellar ages \citep{Pinsonneault2014, Pinsonneault2018, Miglio2020}. At present, thanks to the high-quality spectroscopic and astrometric measurements available, it becomes possible to estimate ages for a large sample of red-giants allowing to analyse larger volumes of the Galactic disc. Further estimations have been made available, through the use of Machine Learning techniques. These techniques use stars with asteroseismic data as a training set \citep[e.g.][]{Ness2016, Mackereth2019, Wu2019, Ciuca2020}.

In the present study, we use the APOGEE atmospheric parameters and abundances \citep{Jonsson2020} and the APOGEE-AstroNN distances, guiding radius and ages \citep{Bovy2019, LeungBovy2019b, Mackereth2019}, contained in the 16$^{th}$ SDSS data release \citep{Ahumada2020}, to probe the radial structure of the Milky Way disc. The outline of the paper is as follows. In Sect.~2, we describe the data and the selection of the different samples. The chemical and age properties of the disc and in particular their evolution with guiding radius are examined in Sect.~3 and 4 respectively. In Sect.~5, we discuss the implications of the new observations for the structure and formation of the Milky Way disc. Finally, Sect.~6 summarises the main results.


\section{Data}
\subsection{APOGEE and APOGEE-AstroNN}
The present study relies on the APOGEE \citep{Jonsson2020} and APOGEE-AstroNN data \citep{Bovy2019} of the 16$^{th}$ SDSS data release \citep{Ahumada2020}. APOGEE DR16 contains 473,307 spectra of 437,445 stars observed from the Apache Point Observatory and from the Las Campanas Observatory. It is the first APOGEE release covering both hemispheres. The APOGEE data include radial velocities, atmospheric parameters (effective temperature, surface gravity and metallicity) and individual abundances for 26 chemical species. Parameters and abundances were derived with the {\it ASPCAP} pipeline \citep{GarciaPerez2016}.

APOGEE-AstroNN is one of the 6 Value Added Catalogues (VAC) associated with APOGEE DR16. APOGEE-AstroNN contains atmospheric parameters, individual abundances for 20 chemical species, distances, ages and kinematic and dynamic parameters. The atmospheric parameters and abundances were re-determined with a Bayesian multi-layer neural network \citep{LeungBovy2019a}, trained on a subset of $\sim$33,000 SN~$>$~200 APOGEE spectra and parameters from the main APOGEE catalogue. APOGEE-AstroNN provides 2 distance estimates \citep{LeungBovy2019b}. The first one was obtained with a neural network designed to determine stellar luminosities from APOGEE spectra, while the second one (the use of which is recommended) is a weighted combination of the first estimate and of Gaia~DR2 parallaxes \citep{GaiaCollaboration2018}. The ages were determined with a third Bayesian neural network \citep{Mackereth2019}, trained on APOKASC-2 \citep{Pinsonneault2018} asteroseismic ages and APOGEE spectra. All three neural networks rely on the open-source Python package {\it astroNN} \citep{LeungBovy2019a}, from which the name of the VAC derives. Finally, APOGEE-AstroNN also includes orbital parameters calculated with the {\it fast} method \citep{MackerethBovy2018} using the MWPotential2014 gravitational potential from \citet{Bovy2015}.

In this study we use the atmospheric parameters, abundances and quality flags from APOGEE\footnote{We used the {\it allStar-r12-l33.fits} version of the APOGEE {\it allStar} file, downloaded on 12/12/2019.}. We complement them with additional information provided by APOGEE-AstroNN\footnote{We used the {\it apogee\_astroNN-DR16-v1.fits} version of APOGEE-AstroNN, downloaded on 27/08/2020.}: i.e. distances, galactocentric cylindrical coordinates (current galactocentric radius $R$, azimuth $\phi$, height $Z$), guiding radius ($R_g$) and, most importantly, ages.

\subsection{Selection of the main sample\label{sect:main}}
\begin{figure}[h!]
   \centering
   \includegraphics[width=\hsize]{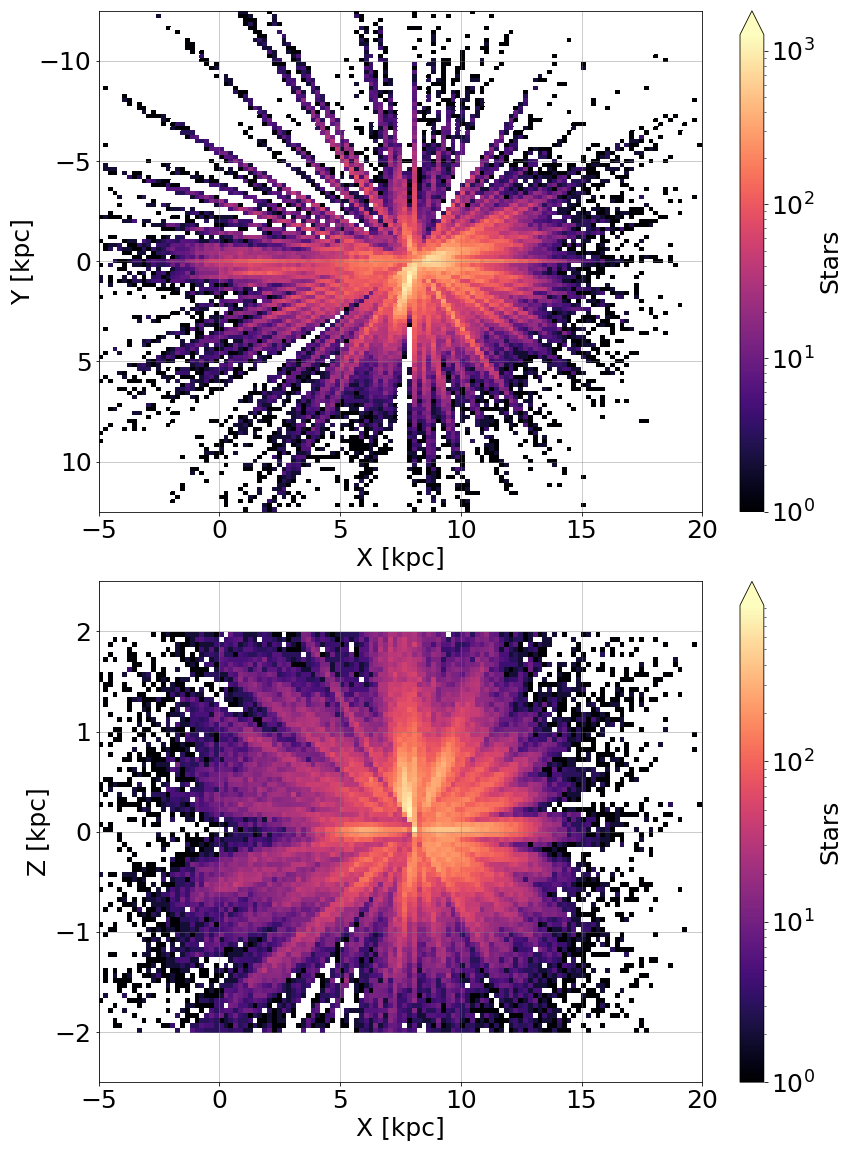}
   \caption{Face-on (top) and edge-one (bottom) distributions of the stars of the main sample. $X$, $Y$ and $Z$ are the  Cartesian Galactic coordinates, with the Z axis positive in the northern Galactic hemisphere. The Galactic centre is located at $X=Y=Z=0$~kpc and the Sun at $X=8.125$~kpc \citep{Gravity2018}, $Y=0$~kpc and $Z=20.8$~pc \citep{BennettBovy2019}. In the top panel, the Galaxy rotates clockwise.}
   \label{fig1}
\end{figure}

Several filters were applied to select the main sample. First, the highest signal-to-noise occurrences of the duplicated APOGEE identifiers were kept, and the others discarded. This cut was made using the {\it EXTRATARG} flag. After this first step, the sample still contained a few thousands records with duplicated GAIA source identifiers. The first occurrence of each was kept and the others were removed. Then, the stars identified by the flags {\it STARFLAG} or {\it ASPCAPFLAG} has having: (i) 40\% or more of bad pixels in their spectra, (ii) a neighbour more than a 100 times brighter, (iii) broad lines or (iv) the {\it STAR\_BAD} bit set to true, were excluded. The stars with a missing or invalid metallicity, magnesium over iron ratio, guiding radius or height were also discarded, as well as the stars with a signal to noise ratio lower than 50 or a metallicity or magnesium over iron uncertainty larger or equal to 0.1~dex. The sample was then restricted to giant stars to avoid potential differential trends in the measure of the parameters of dwarfs and giants and in particular chemical abundances. The stars with effective temperatures and surface gravities respectively in the ranges [3200, 7500]~K and [0.5, 3.5]~dex were selected as giants. The distances which were not at least 5 times larger than their uncertainties were considered as inaccurate and the corresponding stars were suppressed. Finally, the sample was trimmed in guiding radius and height, respectively to $[0, 18]$~kpc and $[-2, 2]$ kpc. In this study, we preferentially use the guiding radius\footnote{For comparison, some of the analysis were repeated using the radius~$R$. They are presented in appendix~\ref{app:D}, \ref{app:G} and \ref{app:H}.} because it averages the effect of blurring and, in this respect, provides a better proxy for the birth radius than the galactic radius itself. Of course, this does not mitigate the effect of churning. Depending on its importance for the Milky Way \citep[see e.g.][]{SchonrichBinney2009a, SchonrichBinney2009b, Minchev2014, Sharma2020}, it is possible that the guiding radii and the birth radii differ for a significant fraction of the stars. The vertical cut seeks to minimise the contamination of the sample by halo stars, without removing too many thick disc stars. After applying all of the above filters, the main sample contains 199,307 stars. Figure~\ref{fig1} shows their distribution in the ($X$, $Y$) and ($R$, $Z$) planes.

\subsection{Separating the thick and thin discs\label{sect:thickthin}}
\begin{figure}[h!]
   \centering
   \includegraphics[width=\hsize]{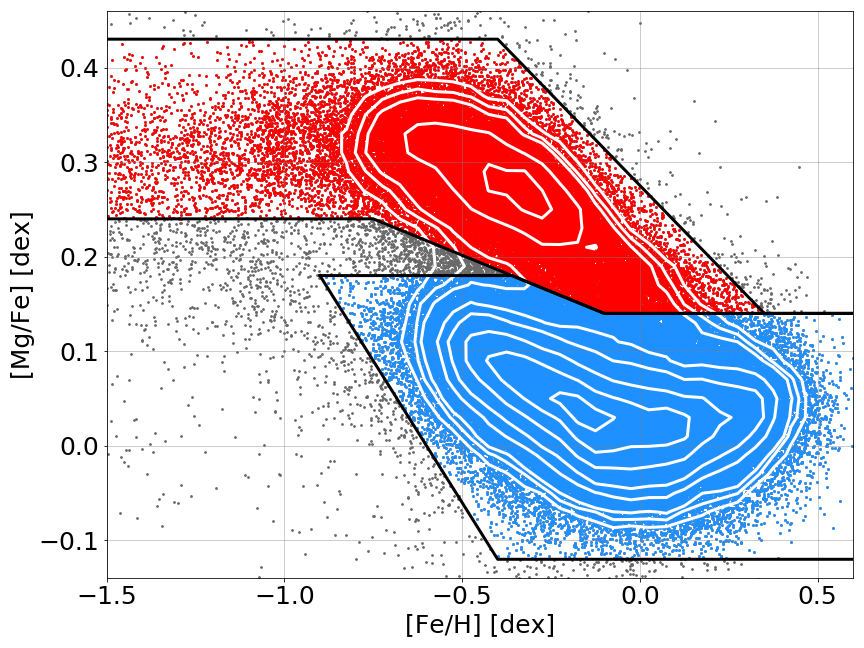}
   \caption{Distribution of the main sample in the ($\FeH$, $\MgFe$) plane. The black segments delineate the thick disc (red dots) and the thin disc (blue dots) selection boundaries. The grey dots are stars attributed to neither populations. The white lines are stellar density iso-contours: i.e.  90\%, 70\%, 50\%, 30\%, 20\%, 10\%, 7.5\% and 5\% of the peak density.}
   \label{fig2}
\end{figure}

In Sect.~\ref{sect:FeHMgFe} and \ref{sect:ageMgFe} we study the disc as a whole. In Sect.~\ref{sect:trend} and \ref{sect:ADF}, we focus on the thin disc only. We use the star locations in the ($\FeH$, $\MgFe$) plane to assign them to the thick or to the thin disc. The boundaries used to separate the two groups are shown in Fig.~\ref{fig2} as black segments. The lower envelope of the thick disc and the upper envelope of the thin disc, were chosen so as to ensure a separation of the high- and low-alpha sequences at all guiding radius (see appendix~\ref{app:A}). The upper envelope of the thick disc and the lower envelope of the thin disc were used to reject outliers, and also for the latter, to further minimise the contamination by the accreted halo sequence \citep{Mackereth2019b, DiMatteo2019}. The stars attributed to the thick and thin disc are shown in Fig.~\ref{fig2}, respectively as red and blue dots. The grey dots, outside of the selection limits, have been associated to neither populations. The white lines are stellar density iso-contours (see caption). The 199,307 main sample stars are distributed into 48,043 thick disc stars, 148,018 thin disc stars and 3,246 "other" stars. In appendix~\ref{app:A}, we show the separation of the thick and thin disc stars as a function of guiding radius.

\subsection{The age sample\label{sect:ageSample}}
\begin{figure}[h!]
   \centering
   \includegraphics[width=\hsize]{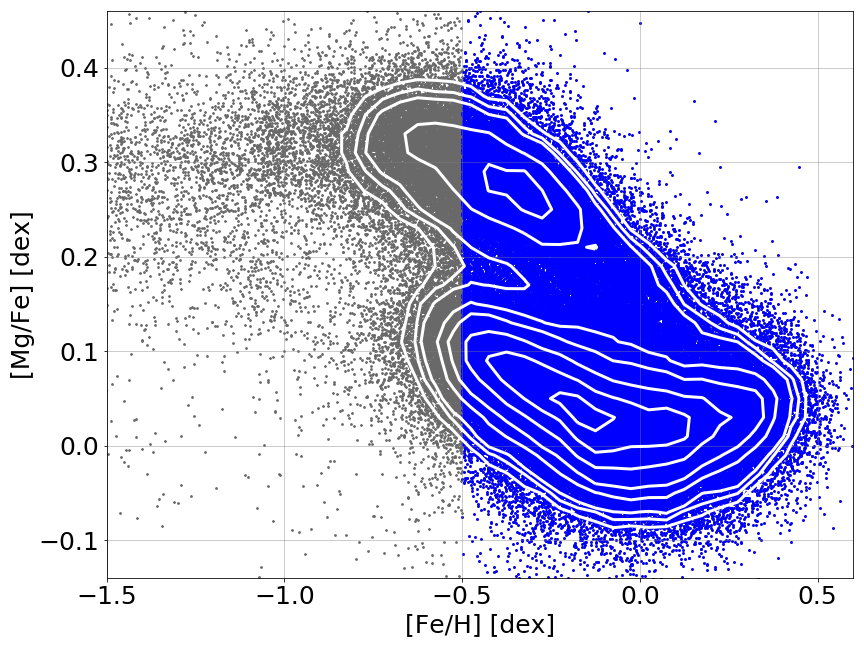}
   \caption{Distribution in the ($\FeH$, $\MgFe$) plane of the main sample stars without valid ages (grey dots) and of the age sample stars (blue dots). The white lines are stellar density iso-contours: i.e.  90\%, 70\%, 50\%, 30\%, 20\%, 10\%, 7.5\% and 5\% of the peak density.}
   \label{fig3}
\end{figure}

\citet{Mackereth2019} and \citet{Bovy2019} recommend to restrict the usage of the APOGEE-AstroNN ages to the stars with metallicities $\FeH > -0.5$~dex. Below this threshold, the APOKASC-2 catalogue \citep{Pinsonneault2018} provides  insufficient coverage to be efficient as training sample for the neural network deriving the ages. When using the ages, we therefore further trim the main sample following the above recommendation. The age sample contains 173,374 stars, of which 31,345 are associated to the thick disc and 141,201 to the thin disc. Fig~\ref{fig3} overplots the main sample (grey dots) and the age sample (blue dots) in the ($\FeH$, $\MgFe$) plane. The metallicity cut truncates the metal-poor part of the high-alpha sequence. For this reason, and also because the oldest ages are affected by systematics (see below), we do a limited use of the ages when studying the high-alpha sequence. The metallicity cut also removes some stars in the low-alpha sequence. As presented in appendix~\ref{app:B}, the percentage of thin disc stars discarded is lower than 5\% up to $R_g = 12$~kpc, is about 12.5\% in the interval $[12, 14]$~kpc and exceeds 25\% beyond $R_g = 14$~kpc. In Sect.~\ref{sect:ages}, we restrict most of the analysis on the thin disc age to $R_g < 14$~kpc and consider with caution the results obtained for the interval $[12, 14]$~kpc.

\citet{Mackereth2019} and \citet{Bovy2019} quote an average precision on the estimates of the age of about 30\%. Moreover, they report a trend in the estimate of the raw ages, the youngest being slightly overestimated, while the oldest are underestimated (slightly at 6-8 Gyr and as much as 3.5 Gyr at 10 Gyr). They explain the underestimation of the old ages by the loss of sensitivity to mass, in the low mass regime, of the carbon and nitrogen features. APOGEE-AstroNN provides three estimates of the age: the raw age determined by the neural network and two ages corrected with two different calibrations of the trend \citep{Bovy2019}. In this study we use one of the two empirically corrected ages, i.e. the {\it AGE\_LOWESS\_CORRECT}, following the recommendation expressed in the documentation.

\begin{figure}[h!]
   \centering
   \includegraphics[width=\hsize]{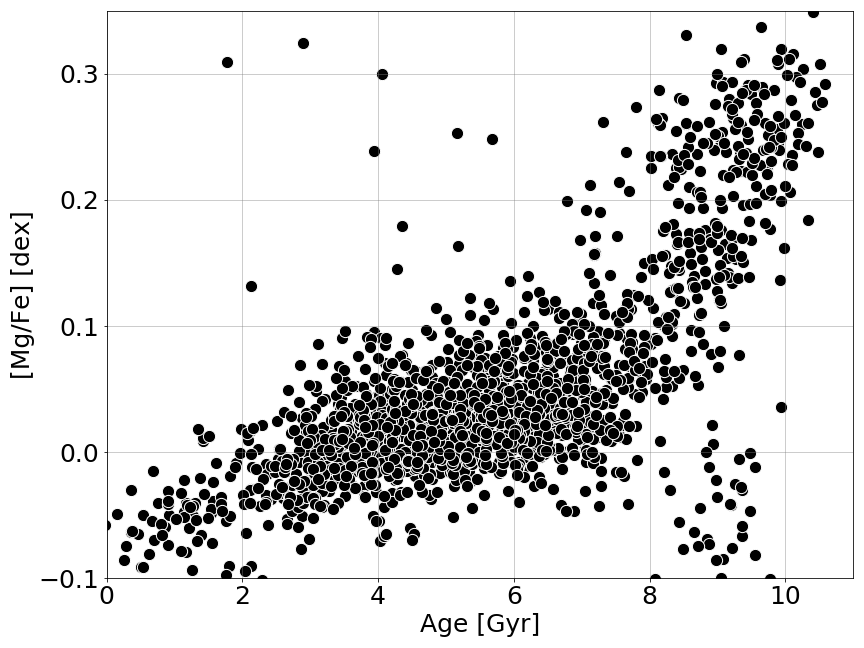}
   \caption{Magnesium over iron ratio as a function of age of the 1,873 age-sample stars located within less than 500~pc from the Sun.}
   \label{fig4}
\end{figure}

Figure~\ref{fig4} shows the $\MgFe$ ratio as a function of age of 1,873 stars from the age sample located at less than 500~pc from the Sun. It can be compared to previous studies of the evolution of the $alpha$-element abundances with age in the solar neighbourhood \citep[e.g.][]{Haywood2013, Bensby2014, Buder2019, DelgadoMena2019, Hayden2020}. The overall behaviour is similar, with two regimes: a smooth increase of the alpha-elements-over-iron ratio in the young and intermediate age stars and a steeper increase in the oldest one. The transition is roughly at the same location: i.e. about 8~Gyr in Fig.~\ref{fig4} and in \citet{Haywood2013} and 8-9~Gyr in \citet{Bensby2014}, \citet{Buder2019} and \citet{DelgadoMena2019}. Looking more closely at the old star sequence in Fig~\ref{fig4}, one can see that it presents a steeper slope than derived in the previous studies and that the oldest ages "saturate" around 11~Gyr. This is consistent with the underestimation of the oldest ages reported by \citet{Mackereth2019} and \citet{Bovy2019}.

\section{Radial structure of the disc: the chemistry\label{sect:chem}}
In this section, we study how the chemical properties of the disc evolve with the guiding radius.

\subsection{Thin disc radial metallicity trend\label{sect:trend}}
\begin{figure}[h!]
   \centering
   \includegraphics[width=\hsize]{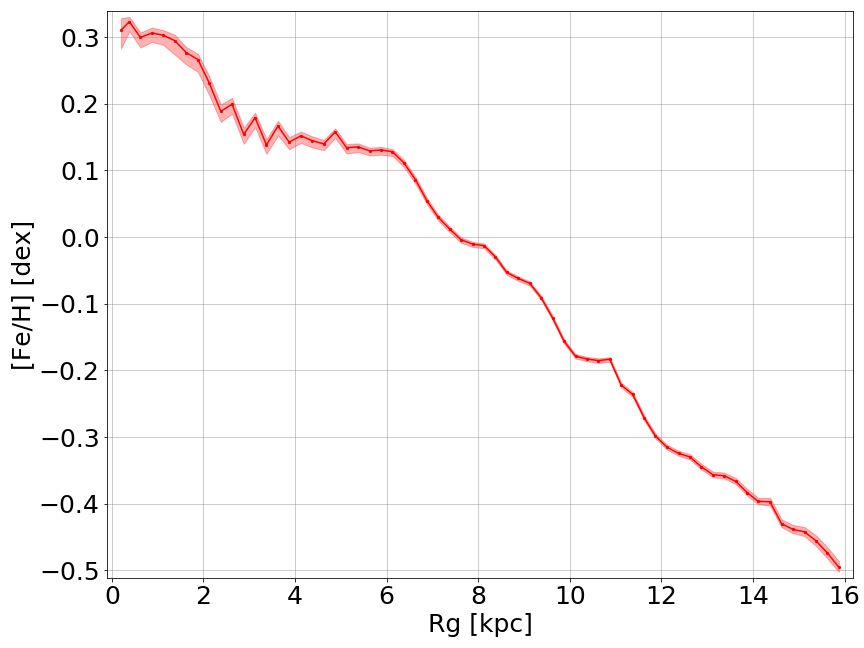}
   \caption{Median metallicity of the thin disc stars as a function of the guiding radius. The median is calculated per bin of 250~pc. The shaded area delimits the $\pm 1 \sigma$ uncertainty on the estimate of the median.}
   \label{fig5}
\end{figure}

We first examine the radial metallicity trend. Figure~\ref{fig5} shows the median metallicity of the thin disc stars (see Sect.~\ref{sect:thickthin}) as a function of the guiding radius. Beyond $Rg \sim 6$~kpc, the thin disc exhibits a negative metallicity gradient, already known and characterised with several tracers, i.e. Cepheids \citep{Luck2011, Lemasle2018}, open clusters \citep{Donor2020, Spina2020} and field stars \citep{Grenon1972, Anders2014, Anders2017, Hayden2015, Haywood2019}. Looking closely at the outer disc, there is a suspicion of a {\it plateau} for $Rg \in [10, 11]$~kpc, which exhibit a slightly more pronounced contrast than the other oscillations along the negative slope. Considered in isolation, this is a relatively weak evidence of a discontinuity in the disc. Yet, we will see in Sect.~\ref{sect:ages} that there are other observables supporting a change in the properties of the disc around $Rg \sim 10$~kpc. As reported by \citet{Hayden2015} and \citet{Haywood2019}, inward of $Rg \sim 6$~kpc, the slope of the metallicity gradient becomes much smoother and the trend becomes almost flat. These two studies, relying respectively on APOGEE DR~12 \citep{Holtzman2015} and DR~14 \citep{Abolfathi2018}, were able to trace the {\it quasi-plateau} up to $R \sim 3.5-4$~kpc. Thanks to the southern hemisphere observations, APOGEE DR16 now provides sufficient statistics to map the trend up to the Galactic centre. Around $Rg \sim 2.5-3$~kpc, the gradient steepens again and reaches a second {\it plateau} in the inner 1-1.5~kpc, around $\FeH \sim 0.3$~dex. Using the APOGEE-AstroNN data, \citet{Bovy2019} mapped the mean metallicity of the Galactic disc up to the Galactic bar. Their map shows that the mean metallicity of the bar is lower than the one of the surrounding disc. As shown in appendix~\ref{app:C}, the apparent differences in the behaviour between a lower metallicity bar and a median thin disc metallicity that rises at $Rg < 3$~kpc is due to the different way we select our respective samples, i.e. we consider only the thin disc stars (defined as the low-alpha stars), while \citet{Bovy2019} used both the high and low-alpha stars.

For comparison, in appendix~\ref{app:D}, we measure the thin disc median metallicity as a function of radius instead of guiding radius. The overall shape of the metallicity trend is very similar to the one derived with the guiding radius: i.e. 2 {\it plateaus} respectively at $R < 2$~kpc and $R \in [4, 6]$~kpc and 2 areas where the metallicity decreases with increasing radius respectively at $R \in [2, 4]$~kpc and $R > 6$~kpc. It could be noted that beyond 6~kpc, the metallicity gradient presents less oscillations as a functions of radius than as a function of guiding radius. This could indicate a dynamical origin of these oscillations. For example, \citet{Khoperskov2020b} showed that the guiding radius space can reveal dynamical resonances which, in direct space, are blurred and therefore more difficult to identify.

\subsection{Patterns and ridge lines in the ($\FeH$, $\MgFe$) plane\label{sect:FeHMgFe}}
\begin{figure}[h!]
   \centering
   \includegraphics[width=\hsize]{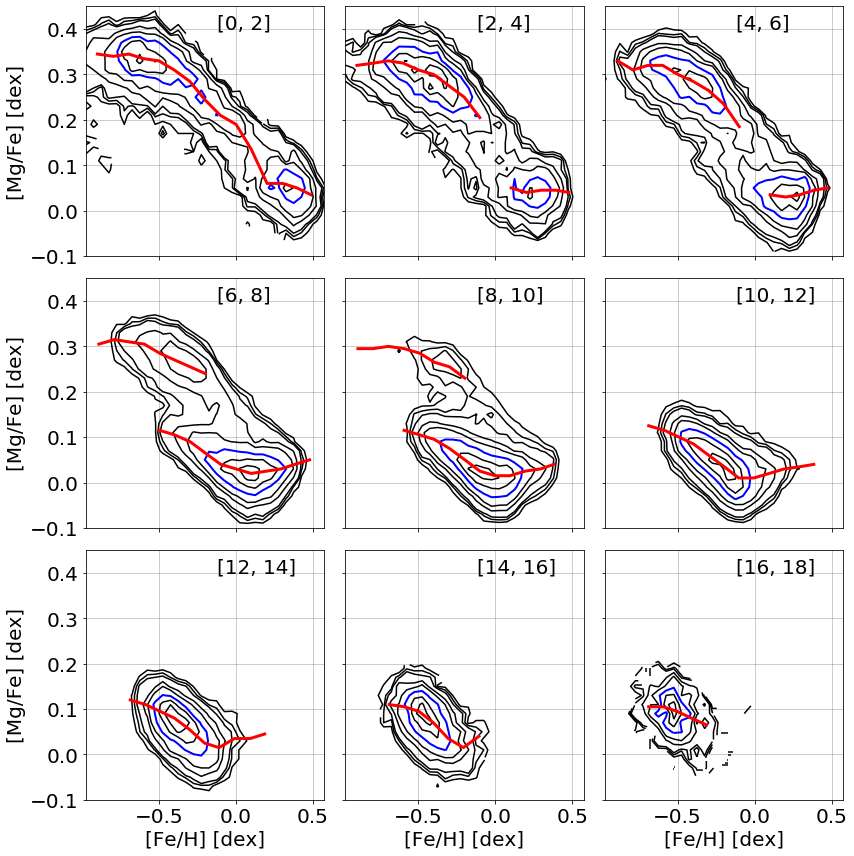}
   \caption{Evolution of the distribution of main sample stars in the ($\FeH$, $\MgFe$) plane as a function of guiding radius, from $Rg \in  [0, 2[$~kpc (top left) to $[16, 18[$~kpc (bottom right). The iso-density contours are plotted at 5, 7.5, 10, 20, 30, 70, 90 (black lines) and 50\% (blue line) of the maximum density. The ridge lines are plotted in red.}
   \label{fig:fehMosaic}
\end{figure}

We now turn our attention to the evolution of the metallicity versus alpha-elements patterns, in the Galactic disc. Figure~\ref{fig:fehMosaic} shows the distribution of the main-sample stars in the ($\FeH$, $\MgFe$) plane. Each thumbnail corresponds to an interval of 2~kpc in guiding radius, from $Rg \in [0, 2]$~kpc (top left) to $[16, 18]$~kpc (bottom right). In each one, iso-density contours are plotted at 5, 7.5, 10, 20, 30, 70, 90 (black lines) and 50\% (blue line) of the maximum density (in the given thumbnail). The red lines are the lines-of-modes of the $\MgFe$ distributions, hereafter referred to as the {\it ridge lines}. They are derived in 2 steps. First, in each $Rg$ interval, the data are divided in metallicity slices of $\Delta \FeH = 0.1$~dex. Then, for each slice, the location of the mode of the $\MgFe$ distribution (or 2 first modes if the distribution is bimodal) is measured. The mode is not a good estimator for the ridge of a sequence which is too steep. For this reason, the steepest part of the high-alpha sequences has been excluded from the analysis. The mode has been preferred to the median, because it allows us to describe the main tracks of multi-sequence data, without the need to separate these sequences first.

Over the last decade, the radial and vertical dependences of the metallicity-alpha-elements distribution have been studied in more and more detail with increasingly larger samples \citep[e.g.][]{Bensby2011, Anders2014, Nidever2014, Hayden2015, Queiroz2020}. Figure~\ref{fig:fehMosaic} is mostly consistent with similar plots showed in the above papers. In the inner 10~kpc it displays two over-densities, a high alpha-element (here $\MgFe$) and a low one. Between $Rg = 6$ and 10~kpc, the two over-densities define two different sequences. In appendix~\ref{app:E}, we note that when the sample is restricted to a $\pm 500$~pc layer around the Galactic plane, two close but separated sequences are observed in the interval $R_g \in [4, 6]$~kpc. Because of their scale-height \citep{Bovy2012}, kinematics \citep{Bensby2003} and age properties \citep{Haywood2013}, these two sequences have been associated to the thick disc (high-alpha) and thin disc (low-alpha) respectively. Moving inward of $R_g = 4-6$~kpc, Fig.~\ref{fig:fehMosaic} shows that the two over-densities connect, through a zone of lower density, to form a single sequence. This is in agreement with the observations of \citet{Hayden2015, Bensby2017, Zasowski2019, Bovy2019, Lian2020c, Lian2020d} who also report a single sequence in the inner disc and/or in the bulge/bar area. Conversely, \citet{RojasArriagada2019} and \citet{Queiroz2020} observe two sequences in the inner regions. In appendix~\ref{app:F}, we compare the distributions of different APOGEE DR16 alpha-elements in the ($\FeH$, $[\alpha/Fe$]) plane (restricting the sample to the stars contained in the interval $R_g \in [0, 2]$~kpc). The different elements produce different patterns: the global alpha-elements abundance\footnote{derived by global fit of the full spectrum \citep{Jonsson2020}.} and oxygen show a double sequence, while magnesium, silicon and calcium present a single sequence. This could explain, at least partly, why \citet{Queiroz2020}, who use a combined $alpha$-elements abundance, observe a double sequence, while we see a single one with magnesium. Yet, this does not explain the discrepancy with \citet{RojasArriagada2019} who also uses magnesium. Beyond $Rg = 10$~kpc, the high-alpha sequence gradually vanishes. This is in agreement with the finding that the thick disc has a shorter scale length than the thin disc \citep{Bensby2011, Cheng2012, Bovy2012}.

\begin{figure}[h!]
   \centering
   \includegraphics[width=\hsize]{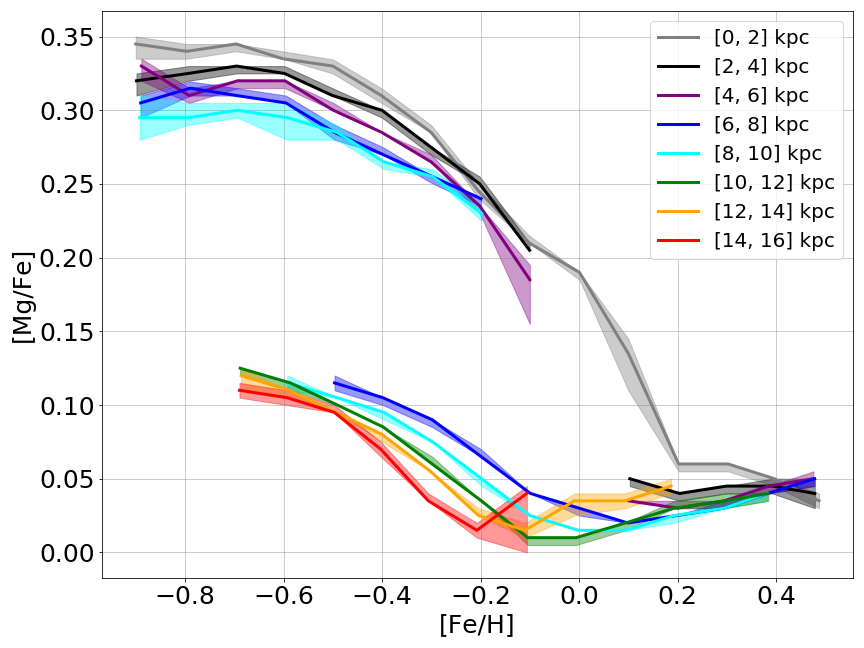}
   \caption{Ridge lines of the ($\FeH$, $\MgFe$) sequences for different $Rg$ annulus, from $[0, 2]$~kpc (gray) to $[14, 16]$~kpc (red). The shaded areas delimit the $\pm 1\sigma$ uncertainties (derived by bootstrap) on the estimates of the modes.}
   \label{fig:fehRidge}
\end{figure}

Although the detailed properties of the inner disc are still debated, the global behaviour of the disc is now rather well established. Here we wish to address a specific question: i.e. the evolution as a function of guiding radius of the location of the sequences in the ($\FeH, \MgFe$) plane. The sequences are broad and significantly overlap from one guiding radius annulus to the next. We therefore use the modes of the $\MgFe$ profiles to describe the {\it ridge line} of the sequences. They are plotted separately per 2~kpc $Rg$ annulus in Fig~\ref{fig:fehMosaic} (as red lines) and together in Fig.~\ref{fig:fehRidge}, allowing us to compare their relative locations. In this last figure, we observe that the thick disc sequences are closely grouped. However, they present a small drift in $\MgFe$ with guiding radius, the innermost sequence ($Rg \in [0, 2]$~kpc) being located a few hundredths of a dex above the outermost sequence ($Rg \in [8, 10]$~kpc). As shown in appendix~\ref{app:E}, the drift remains essentially the same when the sample is sliced by interval of distance to the Galactic plane. The drift is observationally consistent with the measurements of \citet{Bensby2017} who noticed that the bulge stars coincide with the upper envelop of the local thick disc stars. \citet{Lian2020d} recently studied the median $\MgFe$ vs metallicity trend\footnote{which is analogous to the ridge line.} in the bulge and compared the behaviours of on-bar and off-bar stars. They observed that the trends separate between
$\FeH \sim -0.3/-0.2$~dex and solar metallicity, the on-bar one passing roughly 0.05-0.1~dex above the off-bar one.
In Fig.~\ref{fig:fehRidge}, the thin disc sequences behave differently at sub-solar and super-solar metallicity. On the metal-rich side, the thin disc sequences between $R_g = 2$ and 12~kpc, converge and overlap. Conversely, on the metal-poor side, the sequences between $Rg = 6$ and 16~kpc, follow parallel paths, shifting to lower metallicity at larger guiding radius. This confirm the prediction for the outer disc of \citet{Haywood2019} who schematized a similar behaviour in the top panel of their Fig.~6. This is also in qualitative agreement with the trend observed by \citet{Ciuca2020} between the inner, local and outer discs (see the left panel of their Fig.~9). These parallel paths are also reminiscent of the parallel chemical tracks in the model of \citet{Sharma2020} (see their Fig.~10). We discuss possible interpretations of this pattern in Sect.~\ref{sect:oneModel}. For comparison, in appendix~\ref{app:G}, we present the ridge lines derived per interval of Galactic radius.

\begin{figure}[h!]
   \centering
   \includegraphics[width=\hsize]{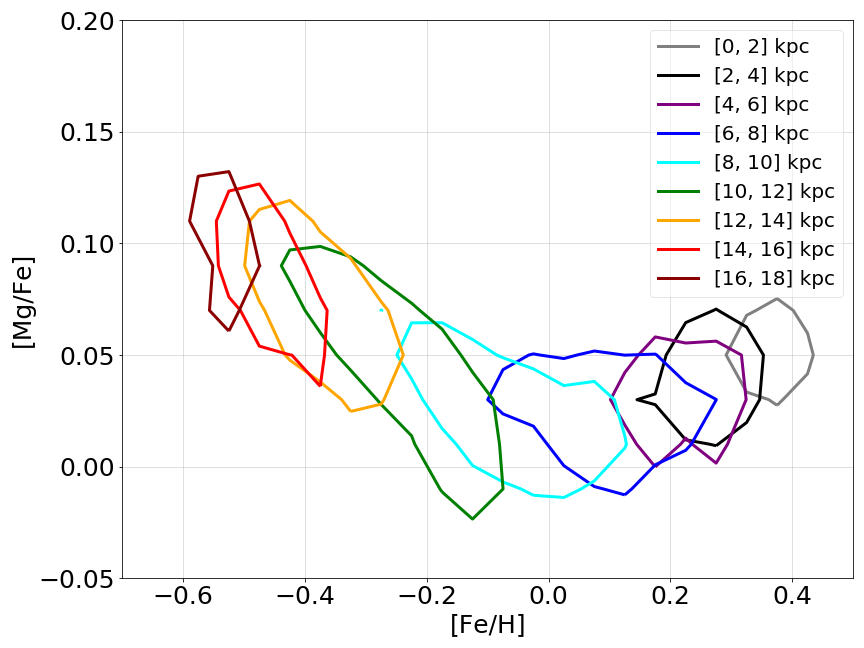}
   \caption{Iso-density contours in the ($\FeH$, $\MgFe$) plane at 70\% of the maximum density, for the thin disc sample. Each contour corresponds to a 2~kpc guiding radius annulus, from $Rg \in [0, 2]$~kpc (gray) to $Rg \in [16, 18]$~kpc (dark red).}
   \label{fig:fehContour}
\end{figure}

The ridge lines do not show where the cores of the star distributions are located along the sequences. To answer this question, Fig.~\ref{fig:fehContour} shows the ($\FeH$, $\MgFe$) iso-density contours at 70\% of the maximum density, for the thin disc sample (as defined in Sect.~\ref{sect:thickthin}), divided in 2~kpc guiding radius annulus, from $Rg \in [0, 2]$~kpc (gray) to $Rg \in [16, 18]$~kpc (dark red). The iso-contours shift towards lower metallicity when the guiding radius increases, with a reduced shift between $[2, 4]$ and $[4, 6]$~kpc. This is, fortunately, consistent with the radial metallicity trend discussed in Sect.\ref{sect:trend}. In the inner disc, the contours present a weak negative $\MgFe$ gradient with guiding radius, which change sign at $Rg \in [8, 10]$~kpc, to become positive, but also steeper in the outer disc.

\section{Radial structure of the disc: the ages\label{sect:ages}}
We are now using the ages of the APOGEE-AstroNN catalogue to study the evolution with guiding radius of the Age-$\MgFe$ distribution (Sect.~\ref{sect:ageMgFe}) and of the age distribution function (Sect.~\ref{sect:ADF})

\subsection{Patterns and ridge lines in the (Age, $\MgFe$) plane\label{sect:ageMgFe}}
\begin{figure}[h!]
   \centering
   \includegraphics[width=\hsize]{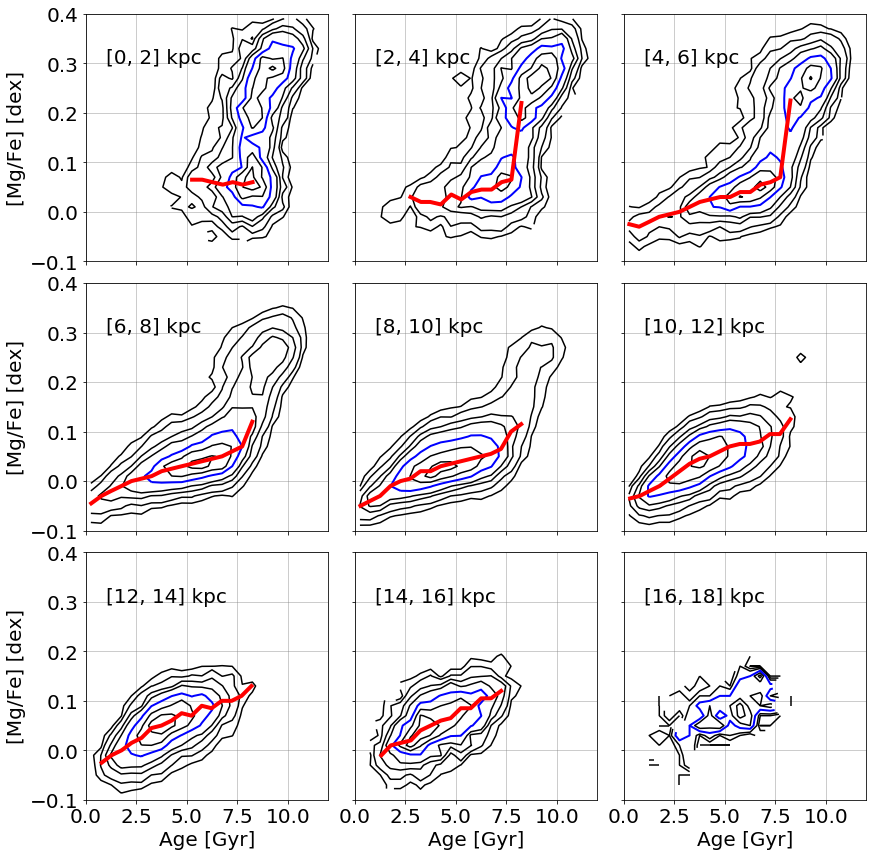}
   \caption{Distributions of the main sample stars in the age-$\MgFe$ plane, from $Rg \in [0, 2]$~kpc (top left) to $[16, 18]$~kpc (bottom right). The iso-density contours are plotted at 5, 10, 20, 30, 70, 90 (black lines) and 50\% (blue line) of the maximum density. The ridge lines, derived over the range $[0, 8.5]$~Gyr, are plotted in red.}
   \label{fig:ageMosaic}
\end{figure}

Fig~\ref{fig:ageMosaic} shows the distribution of the age-sample stars in the age-$\MgFe$ plane. The sample is divided with the same bins in guiding radius as before. In each panel, the iso-density contours are plotted at 5, 10, 20, 30, 70, 90 (black) and 50\% (blue) of the maximum density. As in section~\ref{sect:FeHMgFe}, the ridge-lines (in red) have been derived using the mode of the $\MgFe$ distributions per interval of $\Delta {\rm age} = 0.5$~Gyr, from 0 to 8.5~Gyr. The ridge-lines were not computed for older stars, because beyond $\sim 9$~Gyr, the ages are starting to be underestimated (see Sect.~\ref{sect:ageSample}, \citet{Mackereth2019} and \citet{Bovy2019}). The 380 stars contained in the outermost annulus, i.e. $Rg \in [16, 18]$~kpc, were insufficient to derive its ridge line.

As shown in Fig~\ref{fig:ageMosaic}, from $Rg = 2$ to 10~kpc, the distribution of the stars in the (age, $\MgFe$) plane presents 2 regimes. First, a low-$\MgFe$ sequence made of young and intermediate age stars and presenting a weak $\MgFe$ gradient. Then, a high-$\MgFe$ sequence exhibiting a steeper gradient. The transition between the 2 regimes occurs around 8~Gyr. The separation into high- and low-$\MgFe$ sequences allows to associate the old-$\MgFe$-rich regime to the thick disc and the younger-$\MgFe$-poor regime to the thin disc. The general pattern is similar to the one observed in the solar neighbourhood \citep{Haywood2013, Bensby2014, Buder2019, DelgadoMena2019, Hayden2020} and at larger distances \citep{Feuillet2019}. The proportion of young stars decreases toward the Galactic centre. Within the innermost 2~kpc,  most stars are older than 5~Gyr, so that the thin disc appears more as an over-density than a full-fledged sequence. As discussed in Sect.~\ref{sect:ageSample}, the under-estimation of the APOGEE-AstroNN ages of stars older than $\sim$9~Gyr most likely leads us to over-estimate the slope of the thick disc sequence, which nonetheless is steeper than the one of the thin disc (as shown in the solar neighbourhood by the above mentioned studies). Beyond $Rg = 10$~kpc, the gradual disappearance of the thick disc makes the sequence of the thin disc the main feature. The majority of its stars are younger than $\sim$8~Gyr.

\begin{figure}[h!]
   \centering
   \includegraphics[width=\hsize]{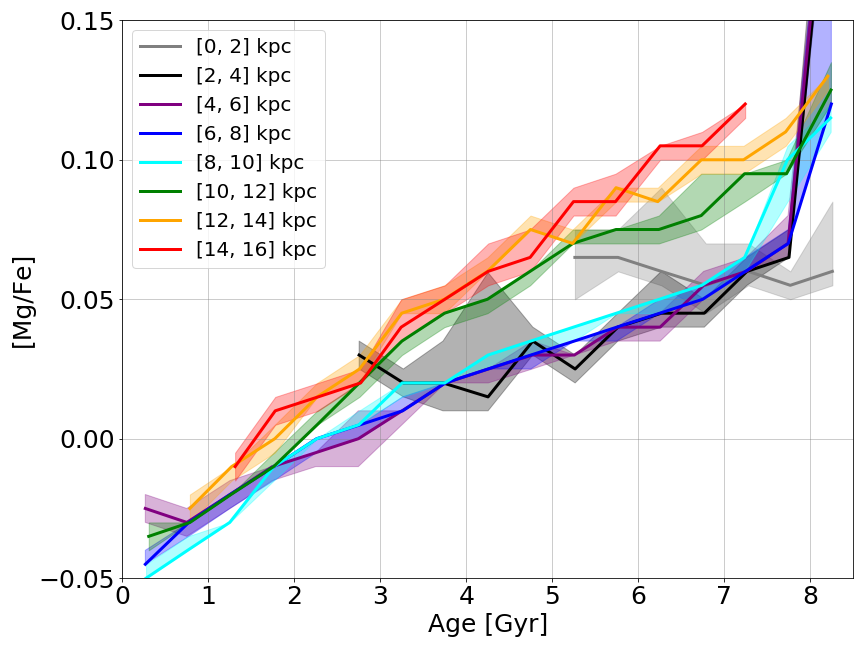}
   \caption{Ridge lines of the age-$\MgFe$ thin disc sequences for different $Rg$ annulus, from $[0, 2]$~kpc (gray) to [14, 16]~kpc (red).}
   \label{fig:ageRidge}
\end{figure}

Figure~\ref{fig:ageRidge} shows the relative locations of the ridge lines of the age-$\MgFe$ thin disc sequences, for different guiding radius. As already seen in the previous figure, the ridge lines present a gradient, the $\MgFe$ ratio increasing with age. Here, we can see that they split in 2 main groups. The ridge lines beyond $R_g = 10$~kpc exhibit a stronger gradient, than the ridge lines below. The two groups overlap below 2~Gyr, and diverge at larger age, producing a gap which reaches $\Delta \MgFe \sim 0.05$~dex at 6~Gyr. In the inner group, the ridge lines from $Rg \in [2, 4]$ to $[8, 10]$~kpc are very tight. The outer group is more loose, i.e. the $Rg \in [12, 16]$~kpc ridge lines appear to be slightly offset above the $Rg \in [10, 12]$~kpc one. This global behaviour is broadly similar to the prediction of \citet{Haywood2019} (see their Figure~6 bottom panel), as well as to the  observations of \citet{Ciuca2020} (see their Figure~9 right panel). Yet, we observe no offset between the inner and local discs, nor a smooth continuous radial trend, but rather a jump between $Rg \in [8,10]$ and $[10, 12]$~kpc potentially followed by a weak outward gradient. For comparison, in appendix~\ref{app:H}, we present the ridge lines derived per interval of Galactic radius.

\subsection{Radial trend of the thin disc age distribution function\label{sect:ADF}}
\begin{figure}[h!]
   \centering
   \includegraphics[width=\hsize]{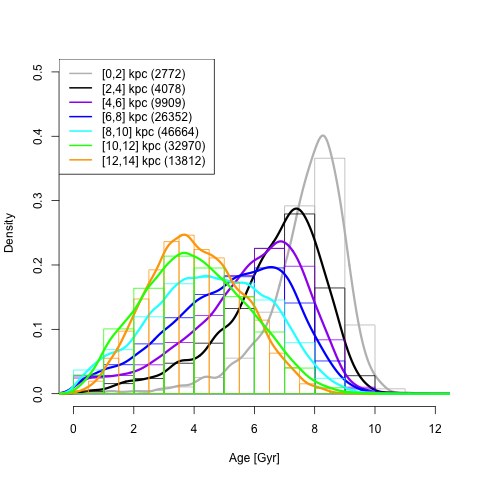}
   \caption{Age distribution functions of the thin disc sample divided in 2~kpc annulus, from $Rg = [0, 2]$ (gray) to $[14, 16]$~kpc (red). The number of stars in each annulus is given in the legend.}
   \label{fig:ADF}
\end{figure}

A key question to understand the assembly of the disc is to probe its star formation history. Fig~\ref{fig:ADF} presents the Age Distribution Functions (ADFs) of the thin disc sample\footnote{Selected as defined in Sect.~\ref{sect:thickthin}.}. From the Galactic centre up to $Rg \in [10, 12$]~kpc, the thin disc ADF shows a radial gradient, i.e. stars are older on average in the central regions, with the onset and conclusion of star formation occurring earlier towards the Galactic centre. The gradient provides a clear signature of the inside-out formation of the thin disc up to $Rg \sim 10-12$~kpc. In contrast, the ADFs of the $R_g \in [10, 12]$ and $[12, 14]$~kpc intervals present very similar profiles, with no further sign of a drift toward younger ages between the two $R_g$ bins. Another, interesting feature is the change in the skewness of the profiles. For $R_g < 8$~kpc, the ADFs are skewed toward young ages, while for $R_g > 10$~kpc the ADFs are skewed toward old ages. It is interesting to note that \citet{Hayden2015} reports a similar evolution of the skewness of the Metallicity Distribution Function (MDF) of the disc stars contained in a $\pm 500$~pc layer around the Galactic plane, the change of sign occurring around $R = 10$~kpc.

The age sample is restricted to stars with $\FeH > -0.5$~dex (see Sect.~\ref{sect:ageSample}). As discussed in appendix~\ref{app:B}, up to $R_g = 12$~kpc, this filter removes less than 5\% of the thin disc stars. This should have a very limited impact on the observed drift of the ADFs with guiding radius as well as on their skewness. The situation is a bit different for the interval $R_g \in [12, 14]$~kpc, where the percentage of stars discarded reaches 12.4\%. The stars removed are by construction the most metal-poor thin disc stars in this range of guiding radius and presumably old stars, which would then preferentially populate the right wing of the profile. Another aspect that should be considered is that our sample is made of giant stars and is therefore not representative of the youngest generations of stars. The profiles of the ADFs, in particular in the first few Gyr, could be missing recently formed non-evolved stars and should be considered with caution.

Numerous models of formation of the disc assume or predict an inside-out formation \citep[e.g.][]{Chiappini1997, Brook2012, Bird2013, Minchev2014, Grand2018, Khoperskov2021, VincenzoKobayashi2020}. Figure~\ref{fig:ADF} provides an observational confirmation of the inside-out formation of the thin disc up to $R_g \sim$10-12~kpc and suggest a coeval formation beyond.

\section{Discussion}
\subsection{Formation of the Galactic disc: the {\it Haywood et al} scenario\label{sect:oneModel}}

\begin{figure*}[h!]
   \centering
   \includegraphics[width=\hsize]{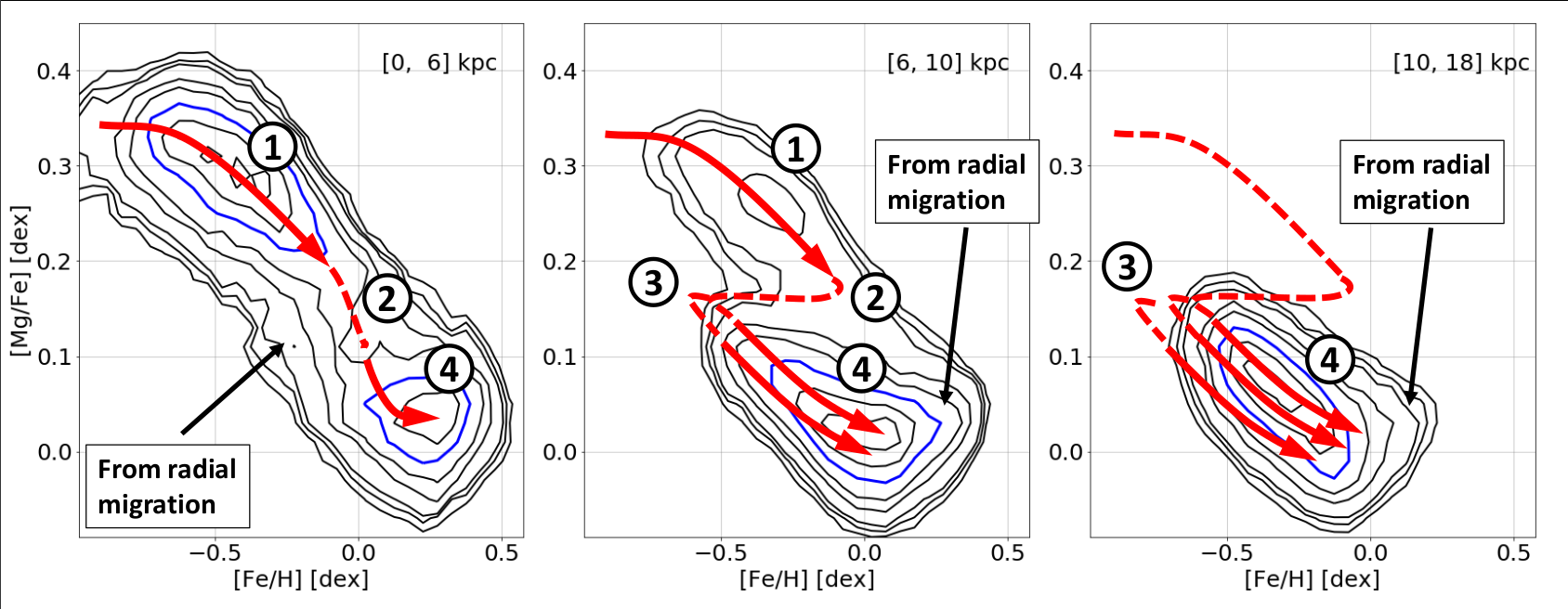}
   \caption{Distribution of the stars of the main sample in the ($\FeH$, $\MgFe$) plane. The stars have been separated in 3 intervals of guiding radius: inner disc, i.e. $Rg \in [0, 6]$~kpc (left), intermediate disc, i.e. $Rg \in [6, 10]$~kpc and outer disc, i.e. $Rg \in [10, 18]$~kpc. The numbered circles refer to the different stages of the formation of the disc (see main text): (1) formation of the thick disc, (2) quenching of the star formation, (3) dilution of the gas in the intermediate and outer thin discs and (4) formation of the thin disc. The red lines are schematic representations of the chemical tracks. The black arrows points to the locus populated by stars which have migrated from another section of the disc.}
   \label{fig:iio}
\end{figure*}

As presented in the introduction, numerous models have been developed to study the formation of the Milky Way disc, and its thick and thin components, by exploring different physical processes, such as: gas-rich mergers \citep{Brook2012, Grand2018, Buck2020, Agertz2020, Renaud2020a, Renaud2020b, Lian2020a, Lian2020b}, radial migration \citep{SchonrichBinney2009a, SchonrichBinney2009b, Loebman2011, Roskar2013, Feuillet2019, Sharma2020, Beraldo2020}, double gas-infall \citep{Chiappini1997, Chiappini2001, Noguchi2018, Spitoni2019, Spitoni2020, Palla2020} or multi-phase star formation history \citep{Lian2020c, Lian2020d}. For several years, our team has been developing a scenario and a model of formation of the Galactic disc \citep{Haywood2013, Snaith2014, Snaith2015, Haywood2015, Haywood2016, Haywood2018, Haywood2019, Snaith2021}, hereafter referred to as the {\it Haywood et al} scenario. In this section, we summarise its main features. Figure~\ref{fig:iio} is intended to help in the presentation of the scenario. It presents the distribution of the main sample in the ($\FeH$, $\MgFe$) plane. The stars have been separated in 3 intervals of guiding radius: inner disc, i.e. $Rg \in [0, 6]$~kpc (left), intermediate disc, i.e. $Rg \in [6, 10]$~kpc (middle) and outer disc (right), i.e. $Rg \in [10, 18]$~kpc. The numbered circles refers to the different stages of the formation of the disc along the chemical tracks which are represented schematically by the red arrows: (1) formation of the thick disc, (2) quenching of the star formation, (3) dilution of the gas and (4) formation of the thin disc.

In this scenario, the disc assembly starts with the thick disc. It forms from a gas-rich turbulent and well-mixed medium during a 3-4~Gyr starburst, about 13 to 9~Gyr ago \citep{Haywood2013, Snaith2014, Snaith2015, Haywood2015}. During the thick disc phase, the star formation extends over approximately 10~kpc radius, but the starburst is energetic enough to pollute the gas at larger distance \citep{Haywood2019}. The efficient mixing of the gas allows the thick disc to evolve along a single chemical track, little dependent on the Galactic radius (label 1 in Fig.~\ref{fig:iio}). About 8~Gyr ago, the star formation quenches \citep{Haywood2013, Snaith2014, Snaith2015, Haywood2016}. This mark the end of the thick disc formation and corresponds in the ($\FeH$, $\MgFe$) plane to an area of lower stellar density (label 2 in Fig.~\ref{fig:iio}). A possible origin\footnote{Other quenching mechanisms have been proposed, such as e.g. a drastic drop of the gas accretion, caused by a switch from cold-flow mode to hot mode \citep{Noguchi2018}, a massive merger, heating the halo gas and inhibiting its accretion \citep{Vincenzo2019} or a depletion of the circumgalactic medium \citep{Kacprzak2020}.} of the quenching is the formation of the Milkay Way bar, which by increasing the velocity dispersion of the gas, could have reduced the star formation by up to a factor 10 \citep{Khoperskov2018}. In the inner disc ($R < 6$~kpc), the star formation resumes from the gas enriched by the thick disc formation and therefore in the continuation of its sequence \citep{Haywood2018}, producing the low-alpha thin disc peak (label 4 in the left panel of Fig.~\ref{fig:iio}). In the intermediate disc ($R \in [6, 10]$~kpc), the metals are diluted by metal-poor gas that may have been in the outer disc for some time already at the time of dilution, but which exact origin is not known (see discussion in \citet{Haywood2019}). The dilution does not affect the inner disc, possibly which is isolated from more external regions by the Outer Lindblad Resonance (OLR) of the {\it recently} formed bar \citep{Haywood2019}. In the intermediate disc, the star formation resumes from the $\MgFe$ ratio produced by the thick disc phase\footnote{magnesium and iron being diluted in similar proportions, the $\MgFe$ abundance ratio is not significantly modified.}, but a lowered metallicity (label 3 Fig.~\ref{fig:iio}). It proceeds at a lower rate than during the thick disc phase. The gas enrichment is dominated by thermonuclear supernovae (SNIa). As the chemical evolution proceeds, the iron content of the ISM increases and its $\MgFe$ abundance ratio decreases, producing the low-alpha thin disc sequence (label 4 in the middle panel of Fig.~\ref{fig:iio}). The gas dilution is stronger at larger distance from the Galactic centre, shifting the chemical tracks to lower metallicity in proportion of their Galactic radius (as schematically represented in Fig.~\ref{fig:iio} by the two parallel arrows). In the outer disc ($R > 10$~kpc), the chemical evolution is similar to the one in the intermediate disc. As the intensity of the dilution continues to increase with the Galactic radius, the outer disc starts to form its stars from a lower metallicity than the intermediate disc. It also starts from a higher $\MgFe$ ratio. A possible explanation for this last point is that the outer disc would have been polluted by the thick disc mostly during the peak of the thick disc starburst phase, which is the period when the energy injected into the ISM reaches its maximum. Once past this peak, the thick disc would have continued its chemical evolution (increasing its metallicity and decreasing its $\MgFe$ ratio) without significantly contaminating the outer disc any more \citep{Haywood2019}.

The thick and thin discs appears as two different overdensities in the ($\FeH$, $\MgFe$) plane. In the {\it Haywood et al} scenario different mechanisms are at work in the inner disc and in the intermediate disc to produce the $\alpha$-bimodality\footnote{Two over-densities belonging to the same sequence.} and the $\alpha$-dichotomy\footnote{Two distinct sequences.} respectively. In the inner disc, the $\alpha$-bimodality is produced by the quenching of the star formation, while in the intermediate disc the $\alpha$-dichotomy is mainly driven by the dilution of the metals before the onset of the thin disc formation.

\subsection{Present work in the context of the {\it Haywood et al} scenario\label{sect:inter}}
In this section, we discuss the observations of Sect.~\ref{sect:chem} and \ref{sect:ages} in the context of the {\it Haywood et al} scenario.

In this scenario, the thick disc is the first component of the disc to form. This is confirmed by Fig.~\ref{fig:ageMosaic}, which shows that the thick disc stars are older than the thin disc stars throughout the disc. In the innermost guiding radius interval ($R_g \in [0, 2]$~kpc), the age difference between the two populations appears small. We suspect that it is, at least partly, an effect of the underestimation of the older ages, which has been shown in sect~\ref{sect:ageSample} to steepen the slope of the thick disc sequence in the solar neighbourhood and very likely has a similar effect here too. The ages used in the present study are not accurate enough in the old stars regime, to assess precisely the period of formation of the thick disc. Yet, Fig.~\ref{fig:ageMosaic} shows that about 8~Gyr ago, the formation of the thick disc was completed.

The {\it Haywood et al} scenario postulates that the thick disc formed from turbulent well-mixed gas and evolved along a chemical track, which depends little on the location in the disc. In Fig.~\ref{fig:fehRidge}, the thick disc ridge lines are indeed closely grouped. They exhibit a weak $\MgFe$ trend, which decreases by about 0.05~dex in $\sim 10$~kpc. This could be the signature of a weak Star Formation Rate (SFR) gradient (stronger in the inner regions than in the outer ones) and/or a weak inside-out formation of the thick disc. Accurate old star ages would help answering this question. Either way, the closeness, smoothness and narrowness of the sequences continue to support the idea that the thick disc formed from a well-mixed medium.

As shown in Fig.~\ref{fig:ADF}, after the quenching, the star formation resumes first in the innermost regions of the disc and then gradually propagates outward. The thin disc forms inside-out up to $R_g \sim 10-12$~kpc. This is a new ingredient for the {\it Haywood et al} scenario, which was not addressing this question so far.

The thin disc ($\FeH$, $\MgFe$) ridge lines present two regimes (Fig.~\ref{fig:fehRidge}). On their metal-poor side and $R_g > 6$ ~kpc, they present parallel sequences shifted in metallicity, while they overlap on their metal-rich side. The metal-poor side is also dominated by stars formed at large guiding radius, while the metal-rich side is mostly populated by inner disc stars (Fig.~\ref{fig:fehContour}). \citet{Haywood2019} predicts such behaviour. At sub-solar metallicity, the parallel ridge lines, all the more metal-poor that they are located far in the disc, are interpreted as the consequence of the increase of the dilution of the gas with increasing radius. The dilution occurred during the quenching phase, setting the initial conditions for the formation of the thin disc, whose evolution then proceeded along parallel chemical tracks. At super-solar metallicity, the ridge-lines raise in $\MgFe$ ratio and overlap with each others. In the scenario discussed here, these stars were born in the inner disc and they did spread at larger radii. The super-solar metallicity stars represent a small fraction of the stars beyond $R_g \sim 6-8$~kpc. Therefore, a moderate radial migration is enough to transport the appropriate number of stars in the intermediate and outer discs \citep{Khoperskov2020}. Similarly a small fraction of solar metallicity stars, born in the intermediate disc, migrates in the inner and outer discs.

Fig.~\ref{fig:appE1} (top panel) shows that in the guiding radius interval $R_g \in [4, 6]$~kpc, within $\pm 500$~pc from the Galactic plane, the thick and thin disc ridge lines are close but disjoint. There are two ways to interpret this in the context of the scenario presented in Sect.~\ref{sect:oneModel}. Some moderate dilution could have occurred inward of 6~kpc and the thin disc ridge line would represent a chemical track. Alternatively, this ridge line could be produced by the contamination by low-alpha sub-solar  metallicity stars born beyond $R_g = 6$~kpc. The bulk of the thin disc stars in that guiding radius interval have super-solar metallicities and are located in the continuation of the thick disc sequence. A moderate radial migration is enough to move the small number of sub-solar metallicity stars required, by a couple of kpc from the intermediate to the inner disc. This favours the contamination hypothesis. It is also supported, by the strong flattening of the thin disc median metallicity gradient around $R_g = 6$~kpc (see Fig.~\ref{fig5}).

Figure~\ref{fig:ageRidge} shows that 7~Gyr ago, the gas in the outer disc ($R_g > 10$~kpc) was enriched in $\MgFe$ by about 0.03-0.05~dex compared to the more internal regions of the disc. As presented in Sect.~\ref{sect:oneModel}, the {\it Haywood et al} scenario explains this offset by different enrichment mechanisms: i.e. in the inner and intermediate discs, the gas was enriched by the full thick disc formation, while in the outer disc it was polluted mostly during the peak of the thick disc starburst. In Fig.~\ref{fig:ageRidge}, the ridge-lines are divided in two relatively narrow set of lines, suggesting a relatively rapid change in the chemical composition of the gas, rather than a smooth evolution with radius. This is not the only property of the disc which changes around $R_g \sim 10$~kpc. The inside-out formation of the disc seems to extend up to $R_g \sim 10-12$~kpc (Fig.~\ref{fig:ADF}). The median metallicity trend presents a {\it plateau} in the $R_g$ interval $[10, 11]$~kpc (Fig.~\ref{fig5}). It is also around $R_g \sim 8-10$~kpc that the skewness of the ADF profiles changes sides (Fig.~\ref{fig:ADF}). Finally, \citet{Hayden2015} reports that the skewness of the MDF changes sign around $R \sim 10$~kpc. We may wonder, if the discontinuity observed in the formation of the Galactic disc between the intermediate and the outer disc could be related to dynamical processes in the disc? We notice that the Outer Lindblad Resonance (OLR) of the bar is placed in the region of transition of both discs. As shown by \citet{Halle2015}, and confirmed by \citet{Khoperskov2020}, the OLR limits the exchange of angular momentum, separating the disc in two distinct parts with limited exchange.

\subsection{Comparison with other models}
In the \citet{Sharma2020} chemodynamical model, the disc $\FeH$ and $\AFe$ content evolves along {\it homologous} chemical tracks, shifted with respect to each other in metallicity in proportion of the star birth radius, i.e. the more metal-rich tracks in the inner regions and the more metal-poor in the outer ones. The tracks are shown in their Fig.~10. The lower half of these tracks (i.e. below $\AFe \sim$~0.1~dex) is in qualitatively good agreement with the ($\FeH, \MgFe$) ridge-lines of Fig.~\ref{fig:fehRidge}. It is in the high-alpha domain that we differ. The {\it knees} of the \citet{Sharma2020} tracks, corresponding to the moment when the thermonuclear super-novae take over the core-collapse super-novae, are shifted to lower metallicities at larger radius and separate from each other. In Fig.~\ref{fig:fehRidge}, the thick disc ridge lines bend around the same metallicity of $\FeH \sim -0.6$~dex at all guiding radius and follow close paths up to $\FeH \sim -0.2$~dex. In the \citet{Sharma2020} model, the double thick/thin disc sequence is produced by the radial migration. In the {\it Haywood et al} scenario, it is the dilution, happening after the thick disc phase, which produces the parallel chemical tracks. This results in the double sequence (mostly visible in the intermediate disc), without the need for a strong radial migration. However, a moderate radial migration transports a small proportion of super-solar metallicity stars from the inner disc to the intermediate and outer discs, possibly caused by a slowing down of the bar pattern speed \citep[see][]{Khoperskov2020}.

In the {\it Haywood} scenario, the inner and intermediate discs are assembled in two distinct star formation phases, separated by a quenching episode. Before the onset of the thin disc formation, the gas is diluted in the intermediate and outer discs. This double-phase process presents similarities with the double infall model, like in the recent work by \citet{Spitoni2019, Spitoni2020}. Indeed, in their model, the second infall dilutes the gas. Yet, the dilution does not occur at the same stage of the chemical evolution. In their model, it occurs around solar $\AFe$ ratio (see Fig.~2 in \citet{Spitoni2019}), while in the {\it Haywood et al} scenario, the gas is diluted {\it earlier}, i.e. when it reaches $\MgFe \sim$~0.1-0.15~dex. As discussed further in the next paragraph, another difference is that in the {\it Haywood et al} scenario, in the inner disc ($R < 6$~kpc) the gas is not diluted and a small proportion of the old super-solar metallicity stars, formed there, migrates to the intermediate and outer discs. The moment when the dilution occurs has of course an influence on the $\FeH$-$\MgFe$ and age-$\MgFe$ tracks followed by the disc along its evolution. In the \citet{Spitoni2020} model, the $\AFe$ ratio reaches a first minimum about 8~Gyr ago, then raises by about 0.1~dex over a brief period of $\sim 2$~Gyr and then smoothly decreases up to the present days to slightly below solar value (see e.g. their Fig.~6 middle panel). This behaviour is a little different from that observed in Fig.~\ref{fig:ageRidge}, where, for a given guiding radius, the $\MgFe$ decreases quasi-linearly with decreasing age. As shown in \citet{Snaith2014}, an earlier dilution combined with a quenching episode produces a smoother age-$\AFe$ trend, more in agreement with Fig.~\ref{fig:ageRidge}.

An important ingredient of the {\it Haywood et al} scenario is that the evolution of the thin disc would not have been monolithic, but would have followed different paths in the inner, intermediate and outer regions. In a recent series of papers, Lian and collaborators also develop this idea. They suggest that the outer disc \citep{Lian2020a}, the inner disc\footnote{that they define as the interval of Galactic radius $R_{GC} \in [4, 8]$~kpc, overlapping partially both with our inner and intermediate discs.} \citep{Lian2020b} and the on-bar and off-bar bulges \citep{Lian2020c, Lian2020d} have formed along different evolutionary channels: double infalls possibly fuelled by a gas-rich merger in the outer and inner discs and a rapid star burst, followed by a quenching episode and then a smooth secular star formation in the bulge\footnote{the off-bar evolution differing from the on-bar one, either by a faster quenching or recent gas accretion.}. \citet{Lian2020a, Lian2020b, Lian2020c} model presents many convergence points with the {\it Haywood et al} scenario: a radial structuration, two different episodes of star formation separated by a quenching, a dilution of the gas in the intermediate and outer discs. Yet, there are also some differences. As in the works of \citet{Spitoni2019, Spitoni2020}, in their model the dilution produced by the second infall happens around slightly super-solar $\MgFe$, while in the {\it Haywood et al} scenario, it happens earlier in the intermediate and outer disc (but not in the inner disc which will form old super-solar metallicity stars, which for a small fraction will migrate outward - see below). The different timing for the dilution changes the way to {\it travel} through the low-alpha sequence in the ($\FeH$, $\MgFe$) plane, i.e. once from meta-poor to metal-rich with the early dilution and twice from metal-rich to metal-poor and then from metal-poor to metal-rich with the later dilution. The motivation for the later dilution in the \citet{Lian2020a, Lian2020b} and in the \citet{Spitoni2019, Spitoni2020} studies is to model with a single chemical track the old super-solar metallicity stars with $\MgFe \in [-0.1, 0.1]$~dex and the younger alpha-poor sub-solar metallicity stars. In the {\it Haywood et al} scenario, the old super-solar metallicity stars with $\MgFe \in [-0.1, 0.1]$~dex are formed in the inner disc along a different evolutionary path and have migrated to the intermediate and, for a small number, to the outer disc. The dilution can then happen at an earlier stage (i.e. at an higher $\MgFe$ ratio), because the intermediate and outer disc tracks do not need to produce super-solar metallicity stars with $\MgFe \in [-0.1, 0.1]$~dex.


\section{Conclusions}
Using a sample of 199,307 giant stars with precise APOGEE abundances and APOGEE-astroNN ages, selected in a $\pm$~2~kpc layer around the Galactic plane, we assessed the radial evolution of the chemical and age properties of the Galactic stellar disc.

The thick disc $\FeH$-$\MgFe$ ridge lines follow closely grouped parallel paths, supporting the idea that the thick disc did form from a well-mixed medium. However, the ridge lines present a small drift in $\MgFe$, which decreases by about $\sim0.05$~dex from $R_g \in [0, 2]$ to $[8, 10]$~kpc. This suggests that the shift observed between the micro-lensed bulge stars and the local thick disc stars \citep{Bensby2017} is not the result from a discrete transition between the disc and the bulge, but rather a smooth radial evolution of the thick disc.

The well documented local and outer thin disc radial metallicity gradient flattens inward of $R_g \sim$~6~kpc, confirming the findings by \citet{Hayden2015} and \citet{Haywood2019}. APOGEE DR16 data allows to trace the {\it plateau} to larger distances than previously accessible. It extends up to $R_g \sim$~2.5-3~kpc, at which point the metallicity raises again.

At sub-solar metallicity, the intermediate and outer thin disc $\FeH$-$\MgFe$ ridge lines follow parallel sequences shifted to lower metallicity as the guiding radius increases. We interpret this pattern, as an indication of a dilution of the inter-stellar medium, which took place after the formation of the thick disc and before the onset of the thin disc formation. At super-solar metallicity, the ridge lines converge and overlap with those of the inner disc. This suggest that the small proportion of super-solar metallicity stars at $R_g > 6$~kpc were probably born in the inner disc and did move outward through radial migration.

The large APOGEE-AstroNN age set allowed us to probe the distributions of stellar ages in the thin disc, as well as the evolution of the age-$\MgFe$ patterns in the disc, from the Galactic centre to $R_g \sim 14$~kpc. An important result provided by this dataset, is that the thin disc presents evidence of an inside-out formation up to $R_g \sim 10-12$~kpc. Moreover, about $\sim 7$~Gyr ago, the outer thin disc ($R_g > 10$~kpc) was more $\MgFe$ enriched than the more internal regions of the disc, by about $\sim 0.03-0.05$~dex. This could be the fossil record of a pollution of the outer disc gas reservoir, by the thick disc during its starburst phase. 

These results support earlier claims \citep[e.g.][]{Haywood2013, Snaith2015, Haywood2019} that the thin disc did form following different chemical paths inward and outward of $R_g \sim 6$~kpc. Several changes or discontinuities are also observed around $R_g \sim 10-11$~kpc. Further work is required to assess their nature: i.e. chemical and/or dynamical. By comparison, the thick disc seems a rather homogeneous structure, from the Galactic centre up to $R_g \sim$~10~kpc. However, ages for the old populations of similar quality to those available for the thin disc would provide precious constrains on the formation of the thick disc and, beyond, to the accreted and/or {\it in-situ} halo.

\begin{acknowledgements}
This study used SDSS IV DR16 APOGEE and APOGEE-AstroNN data. Funding for the Sloan Digital Sky Survey IV has been provided by the Alfred P. Sloan Foundation, the U.S. Department of Energy Office of Science, and the Participating Institutions. SDSS-IV acknowledges support and resources from the Center for High Performance Computing  at the University of Utah. The SDSS website is www.sdss.org.

SDSS-IV is managed by the Astrophysical Research Consortium for the Participating Institutions of the SDSS Collaboration including the Brazilian Participation Group, the Carnegie Institution for Science, Carnegie Mellon University, Center for Astrophysics | Harvard \& Smithsonian, the Chilean Participation Group, the French Participation Group, Instituto de Astrof\'isica de Canarias, The Johns Hopkins University, Kavli Institute for the Physics and Mathematics of the Universe (IPMU) / University of Tokyo, the Korean Participation Group, Lawrence Berkeley National Laboratory, Leibniz Institut f\"ur Astrophysik Potsdam (AIP),  Max-Planck-Institut f\"ur Astronomie (MPIA Heidelberg), Max-Planck-Institut f\"ur Astrophysik (MPA Garching), Max-Planck-Institut f\"ur Extraterrestrische Physik (MPE), National Astronomical Observatories of China, New Mexico State University, New York University, University of Notre Dame, Observat\'ario Nacional / MCTI, The Ohio State University, Pennsylvania State University, Shanghai Astronomical Observatory, United Kingdom Participation Group, Universidad Nacional Aut\'onoma de M\'exico, University of Arizona, University of Colorado Boulder, University of Oxford, University of Portsmouth, University of Utah, University of Virginia, University of Washington, University of Wisconsin, Vanderbilt University, and Yale University.

This research made use of the NASA's Astrophysics Data System (ADS) bibliographic services, as well as of the open-source Python packages {\it Pandas} \citep{McKinney2010}, {\it Numpy} \citep{Harris2020}, {\it Matplotlib} \citep{Hunter2007} and {\it Astropy}\footnote{http://www.astropy.org} \citep{Astropy2013, Astropy2018}.

ONS thanks the DIM ACAV+ for founding during the preparation of this work.

\end{acknowledgements}

%
%

\bibliographystyle{aa}
\bibliography{disc}


\begin{appendix}
%
%
\section{Thick/thin discs separation versus guiding radius \label{app:A}}
\begin{figure}[h!]
   \centering
   \includegraphics[width=\hsize]{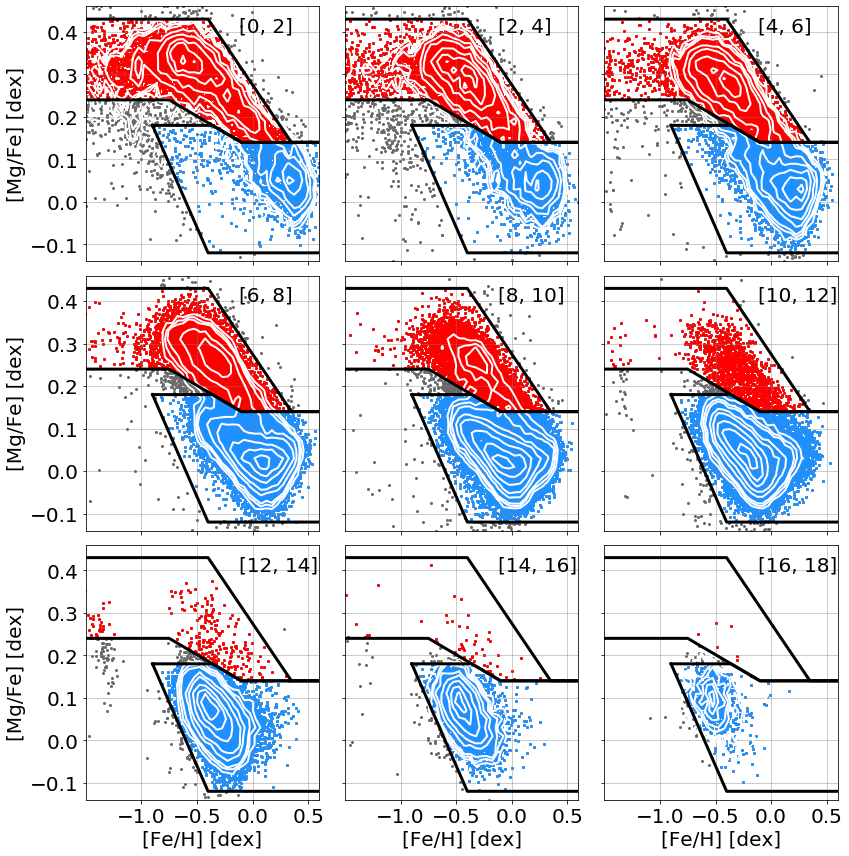}
   \caption{Distribution of the main sample stars in the ($\FeH$, $\MgFe$) plane for different intervals of guiding radius, from $R_g \in [0, 2]$~kpc (top left) to $[16, 18]$~kpc (bottom right). The black segments delineate the thick disc (red dots) and the thin disc (blue dots) boundaries. The grey dots are stars attributed to neither populations. The white lines are stellar density iso-contours drawn at 90\%, 70\%, 50\%, 30\%, 20\%, 10\%, 7.5\% and 5\% of the peak density.}
   \label{fig:appA}
\end{figure}

As described in Sect.~\ref{sect:thickthin}, the thick and thin disc stars were identified based on their location in the ($\FeH$, $[Mg/Fe$) plane. The same criteria were applied for all stars, regardless of their location in the Galactic disc, and this to avoid introducing a selection bias that would be function of radius. Yet, as discussed in Sect.~\ref{sect:FeHMgFe}, the morphology of the high and low alpha sequences changes with guiding radius. The thick disc lower boundary and the thin disc upper boundary were chosen so as to separate the high- and low-alpha overdensities\footnote{We use here the term {\it overdensities} rather than {\it sequences}, because in the inner disc, the ($\FeH$, $\MgFe$) distribution follows a single sequence with two overdensities, a high-alpha and a low-alpha one.} at all guiding radius. Similarly, the thick disc upper envelop and the thin disc lower envelop were defined as to delineate the high and low alpha overdensities without truncating them in any of the guiding radius intervals. Fig.~\ref{fig:appA} illustrates the identification of the thick (red dots) and thin disc stars (blue dots), in 9 successive guiding radius intervals, each 2~kpc wide. The black segments draw the selection limits.

%
%
\section{Age sample versus guiding radius \label{app:B}}
\begin{figure}[h!]
   \centering
   \includegraphics[width=\hsize]{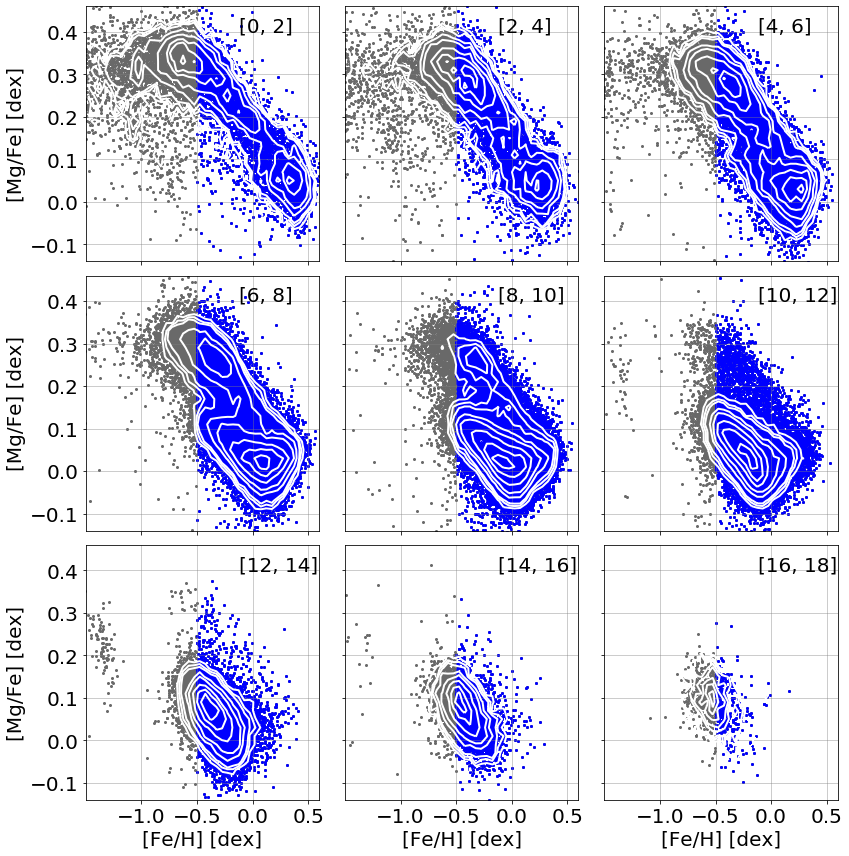}
   \caption{Distribution of the main sample stars in the ($\FeH$, $\MgFe$) plane for different intervals of guiding radius, from $R_g \in [0, 2]$~kpc (top left) to $[16, 18]$~kpc (bottom right). The stars of the age sample are represented as blue dots, while the gray dots are stars from the main sample with invalid ages. The white lines are stellar density iso-contours drawn at 90\%, 70\%, 50\%, 30\%, 20\%, 10\%, 7.5\% and 5\% of the peak density.}
   \label{fig:appB}
\end{figure}

Figure~\ref{fig:appB} shows the distributions of the stars of the age sample (blue dots) and of the main sample with invalid ages (gray dots) in the ($\FeH$, $\MgFe$) plane,  for different intervals of guiding radius, from $R_g \in [0, 2]$~kpc (top left) to $[16, 18]$~kpc (bottom right). As the guiding radius increases, the low-alpha sequence gradually extends toward lower metallicity. As a consequence, the limit at $\FeH > -0.5$~dex, defining the valid ages, excludes gradually a larger fraction of the low-alpha sequence. Table~\ref{tab:appB} provides the percentage of the main sample thin disc stars without valid age as a function of guiding radius. Up to $R_g = 12$~kpc, the metallicity cut removes less than 5\% of the thin disc stars and therefore should have a limited effect on the analysis of the ages presented in Sect.~\ref{sect:ADF}. We should be more cautious with the results obtained in the interval $R_g \in [12, 14]$~kpc, in which the stars more metal-poor than $-0.5$~dex represent 12.4\% of the main sample thin disc stars. Beyond $R_g = 14$~kpc, the thin disc sample was considered as too significantly truncated by the metallicity filter. It was excluded from the analysis carried out in Sect.~\ref{sect:ADF}.

\begin{table}
\caption{Percentage of the main sample thin disc stars without valid age (col. 2, 4 and 6) as a function of guiding radius (col. 1, 3 and 5; expressed in kpc).}
\label{tab:appB}
\centering
\begin{tabular}{c c | c c | c c}
\hline \hline
$R_g$    & Per.      & $R_g$      & Per.      & $R_g$      & Per. \\ \hline
$[0, 2]$ & 2.5\%     & $[6, 8]$   & 0.9\%     & $[12, 14]$ & 12.4\%    \\
$[2, 4]$ & 1.5\%     & $[8, 10]$  & 1.8\%     & $[14, 16]$ & 28.1\%    \\
$[4, 6]$ & 0.9\%     & $[10, 12]$ & 4.7\%     & $[16, 18]$ & 57.3\%    \\
\end{tabular}
\end{table}

%
%

\section{Inner disc metallicity trend\label{app:C}}
\begin{figure*}[h!]
   \centering
   \includegraphics[width=0.33 \hsize]{apogee_08_R_FeH_thinDisc_medianTrend.png}
   \includegraphics[width=0.33 \hsize]{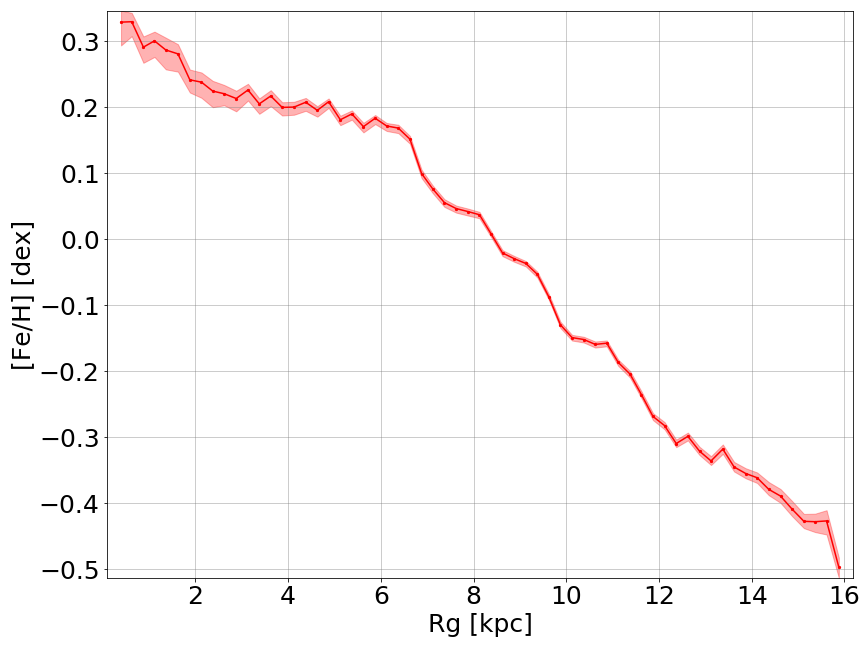}
   \includegraphics[width=0.33 \hsize]{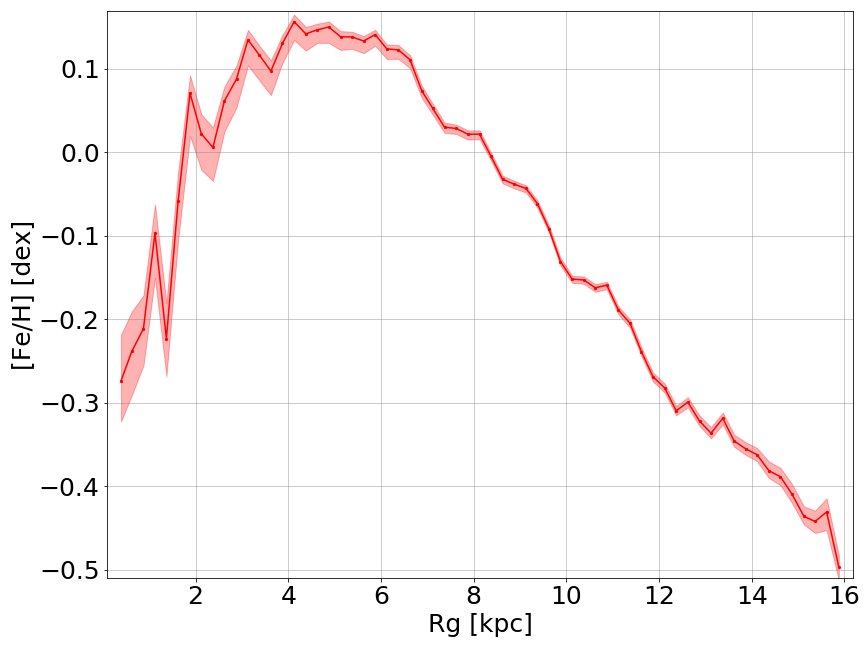}
   \includegraphics[width=0.33 \hsize]{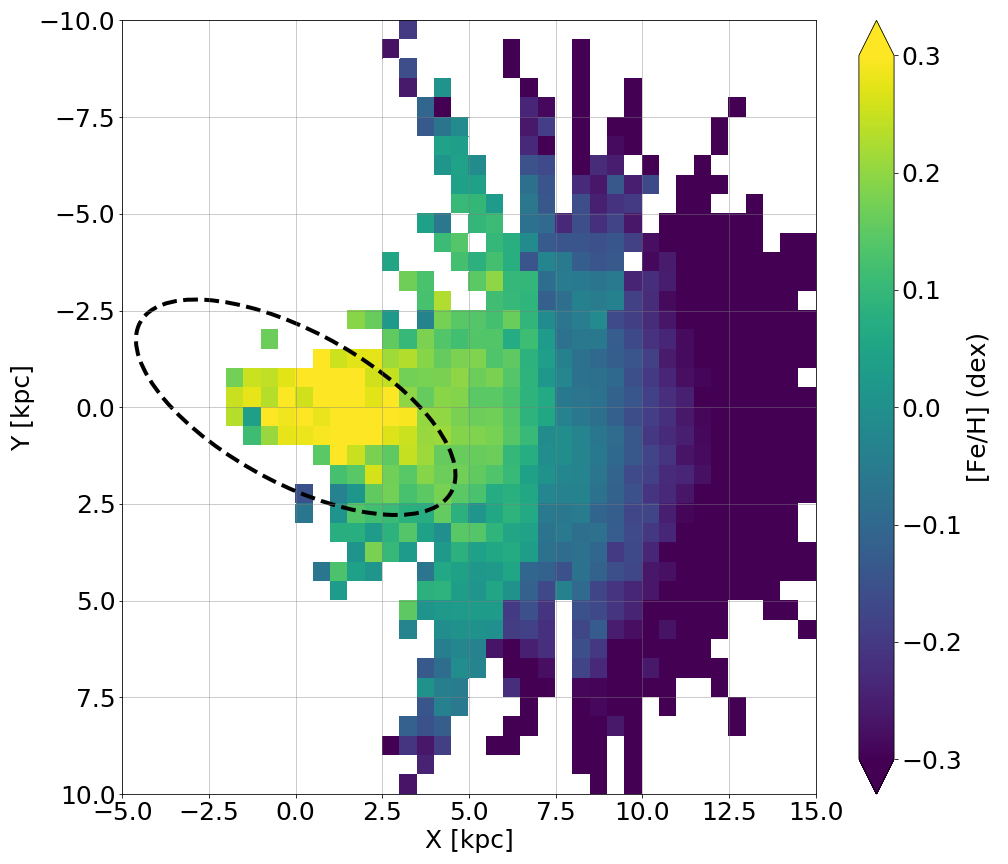}
   \includegraphics[width=0.33 \hsize]{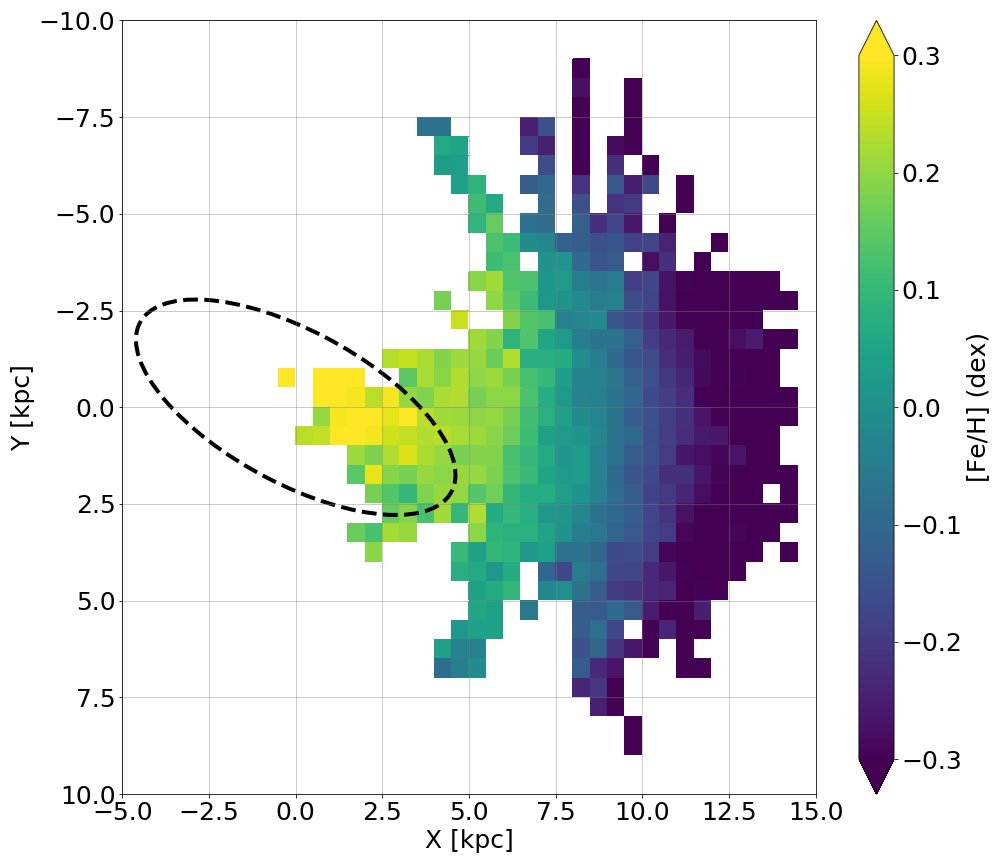}
   \includegraphics[width=0.33 \hsize]{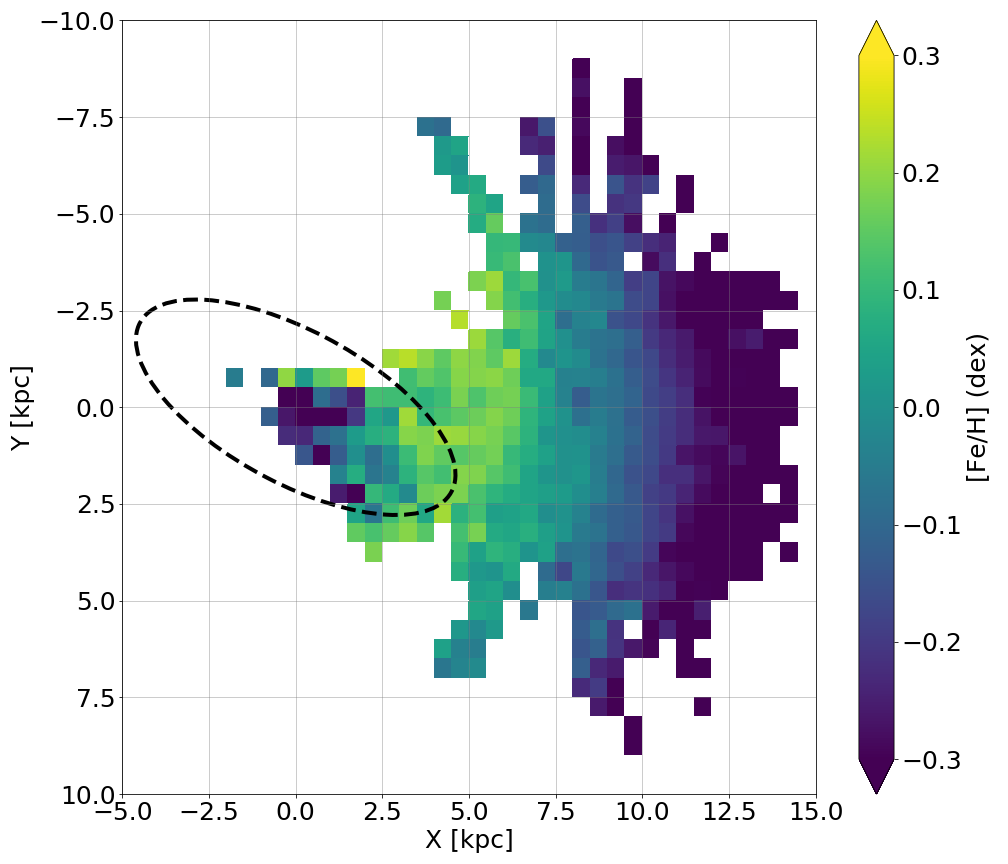}
   \caption{{\bf Top row:} median metallicity trend as a function of guiding radius for three different samples, i.e. the thin disc stars contained in the $\pm2$~kpc layer around the Galactic plane (left), the thin disc stars in the $\pm300$~pc layer (middle) and all the disc stars in the $\pm300$~pc layer (right). The left plot is the same as Fig~\ref{fig5}, reproduced here to facilitate the comparison. The medians are calculated per bin of 250~pc. The shaded area delimits the $\pm1\sigma$ uncertainty on the estimate of the median. {\bf Bottom row:} median metallicity (X-Y) maps for the same three samples as on the top row, i.e. thin disc with $Z \in [-2, 2]$~kpc (left), thin disc with $Z \in [-300, 300]$~pc (middle) and full disc with $Z \in [-300, 300]$~pc (right). X and Y are the Cartesian Galactic coordinates, with the Galactic centre located at $X = Y = 0$~kpc and the Sun at $X = 8.125$~kpc and $Y = 0$~kpc. The Galaxy rotates clockwise. The bar is schematically represented as in \citet{Bovy2019}, by a dashed ellipse with a semi-major axis of 5 kpc and an axis ratio of 0.4, inclined by 25$^\circ$ with respect to the Galactic centre-Sun direction.}
   \label{fig:appC}
\end{figure*}

In Sect.~\ref{sect:trend}, we observe that the median metallicity at $R_g < 2.5-3$~kpc is higher than at larger guiding radius. This could seem in contradiction with the mean metallicity map of \citet{Bovy2019} (see their Fig.~5, middle panel), which shows a bar more metal-poor that the surrounding disc. Yet, to make a proper comparison, one should take into account the way the samples were selected. Our sample is made of thin disc stars (selected as low-alpha stars) contained in a $\pm2$~kpc layer around the Galactic plane. \citet{Bovy2019} used all stars (with no filter on the alpha-elements abundances) contained in a thinner layer of $\pm300$~pc around the Galactic plane. In the top row of Fig.~\ref{fig:appC}, we measure the median metallicity as a function of guiding radius with different samples, applying gradually \citet{Bovy2019}'s selection criteria. The top left panel is a copy of Fig.~\ref{fig5}, shown here to facilitate the comparison. The top middle panel shows the metallicity trend of the thin disc stars, restricted to the $\pm300$~pc layer around the Galactic plane. As with the broader layer, the metallicity rises at $R_g < 2.5-3$~kpc. The top right panel shows the metallicity trend in the same thin $\pm300$~pc layer, but including all disc stars. Consistently, with \citet{Bovy2019} map, when applying similar selection criteria, the median metallicity decreases from $R_g = 3-4$~kpc down to the Galactic centre.

In order to better capture the 2D geometry of the disc metallicity behaviour, in the bottom row of Fig.~\ref{fig:appC} we plot the median metallicity maps of the disc, using the three same samples as on the top row. The left and middle bottom panel (mapping the thin disc stars respectively in the $\pm2$~kpc and $\pm300$~pc layers) both show a bar area, delineated by the dashed ellipse, more metal-rich than the surrounding disc. The right bottom panel is analogous\footnote{\citet{Bovy2019} used APOGEE-AstroNN metallicity, while we are using ASPCAP metallicity. As can be seen from Fig.~\ref{fig:appC} this has no effect on our conclusions.} to the middle panel of \citet{Bovy2019}'s Fig.~5, relying on similar selection criteria. Consistently with their result it shows a bar area, more metal-poor than its surrounding.

The apparent difference in the behaviour of our radial metallicity trend and \citet{Bovy2019} metallicity map, simply reflects a different behaviour between the low-alpha stars and the whole bar-disc stars (high and low alpha together).

%
%
\section{Metallicity trend versus radius\label{app:D}}
\begin{figure}[h!]
   \centering
   \includegraphics[width=\hsize]{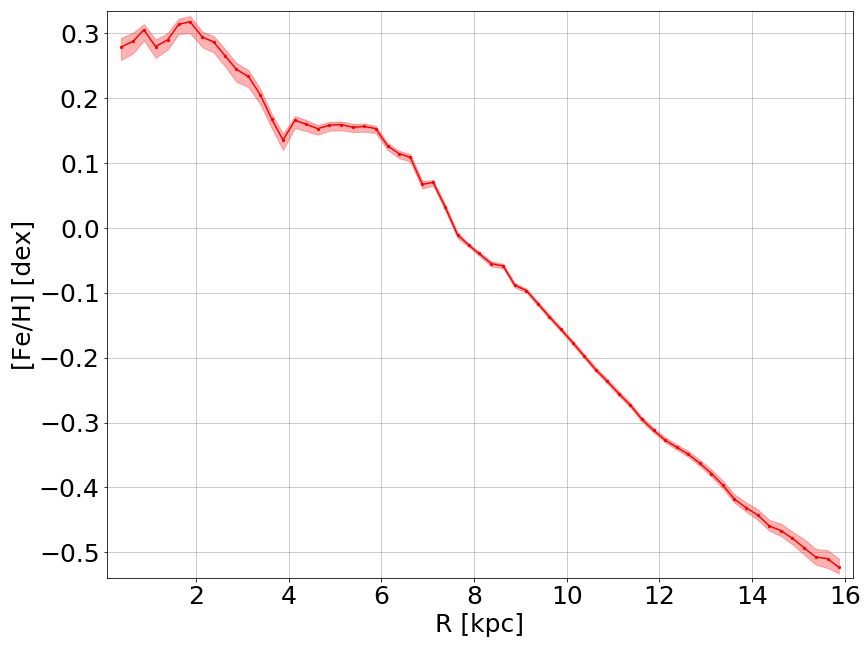}
   \caption{Median metallicity of the thin disc stars as a function of the Galactic radius. The median is calculated per bin of 250~pc. The shaded area delimits the $\pm1\sigma$ uncertainty on the estimate of the median.}
   \label{fig:appD}
\end{figure}

As explained in Sect~\ref{sect:main}, we have chosen to conduct most of our analysis using the guiding radius, with the aim to mitigate the displacements of the stars caused by blurring. It is nonetheless interesting to repeat the analysis with the Galactic radius and to compare the results. Figure~\ref{fig:appD} shows the median metallicity of the thin disc stars as a function of Galactic radius. The overall trend is very similar to the one observed in Fig~\ref{fig5}. In the innermost 2~kpc, the metallicity is roughly constant around $\FeH \sim +0.3$~dex. It then decreases up to $R = 4$~kpc, where it reaches the second {\it plateau} which extends up to $R = 6$~kpc \citep[as previously reported by][]{Hayden2015, Haywood2019}. Beyond starts the well studied disc radial metallicity gradient. Compared to Fig.~\ref{fig5}, the two transitions at $R = 2$ and 4~kpc seems more abrupt and marked. Beyond 6~kpc, the gradient fluctuate less with respect to the Galactic radius than with respect to the guiding radius. The dynamical blurring is a plausible explanation for the smoother metallicity trend observed as a function of Galactic radius.

%
%
\section{Chemical patterns with different vertical cuts\label{app:E}}
\begin{figure}[h!]
   \centering
   \includegraphics[width=0.82 \hsize]{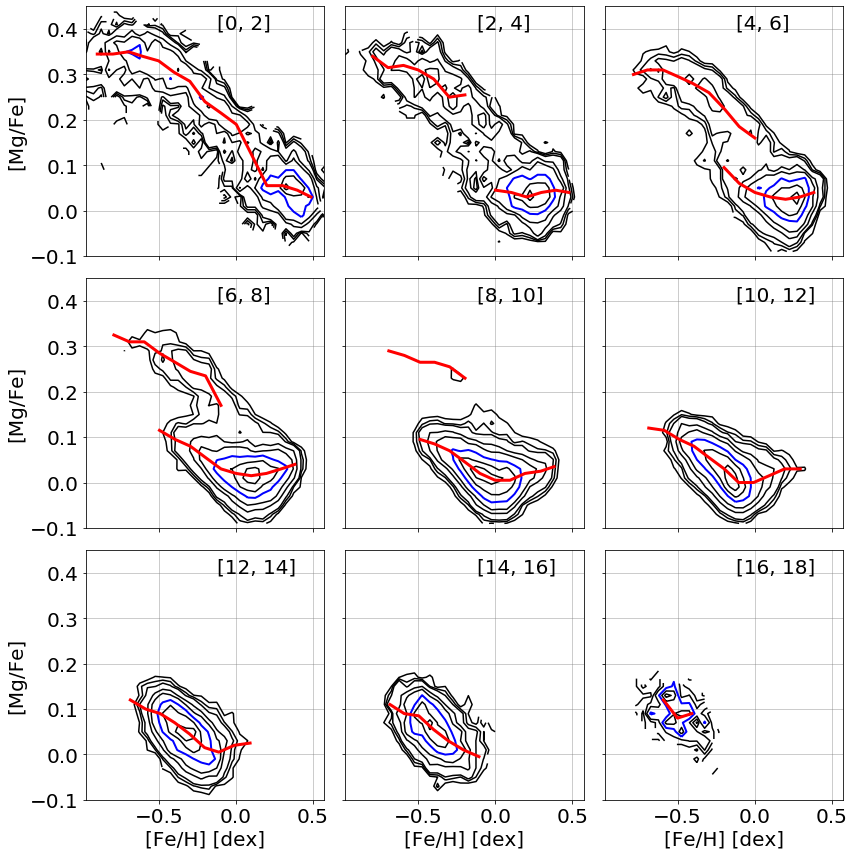}\\
   \vspace{0.2cm}
   \includegraphics[width=0.82 \hsize]{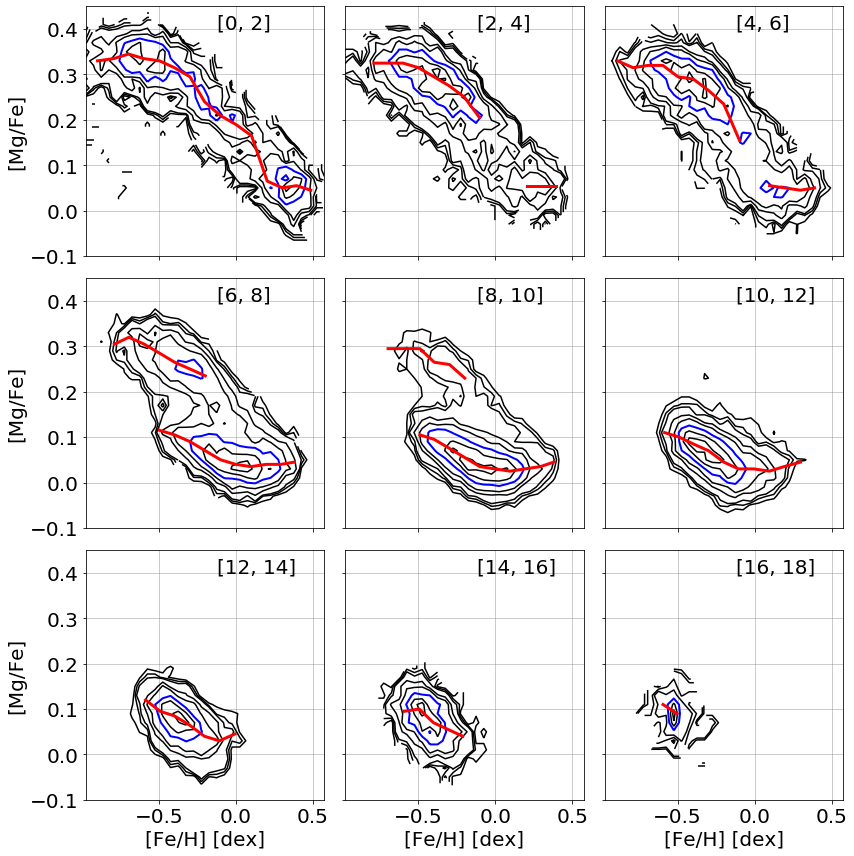}\\
   \vspace{0.2cm}
   \includegraphics[width=0.82 \hsize]{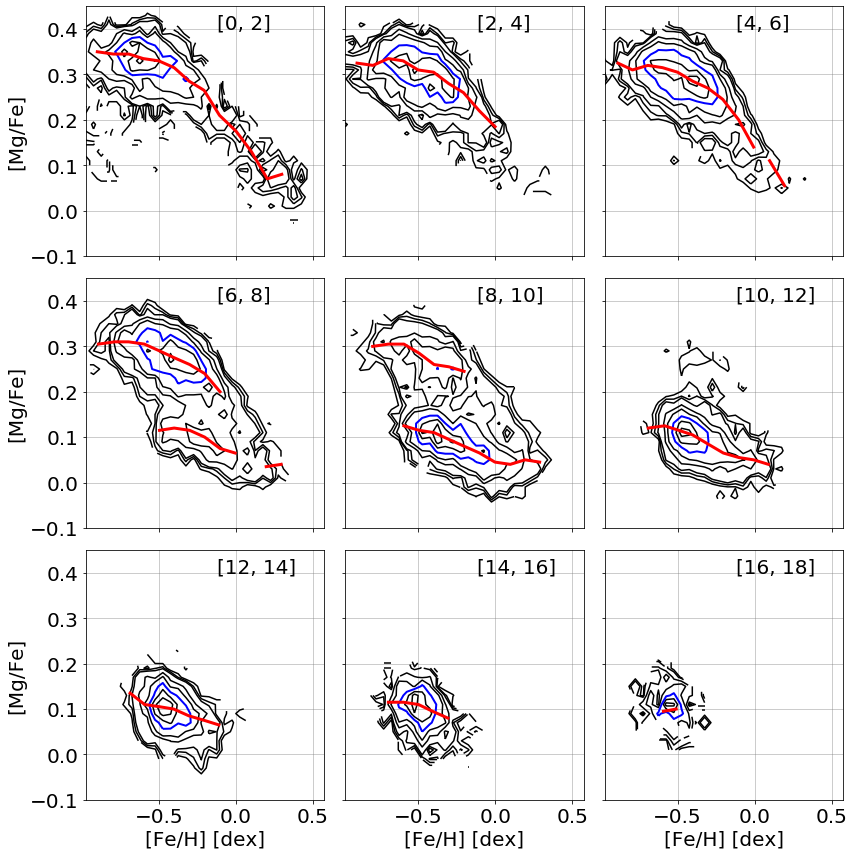}
   \caption{Same as Fig.~\ref{fig:fehMosaic}, for different intervals of distance to the Galactic plane: $|Z| \in [0, 500]$~pc (top), $|Z| \in [500, 1000]$~pc (middle) and $|Z| \in [1000, 2000]$~pc (bottom).}
   \label{fig:appE1}
\end{figure}

\begin{figure}[h!]
   \includegraphics[width=\hsize]{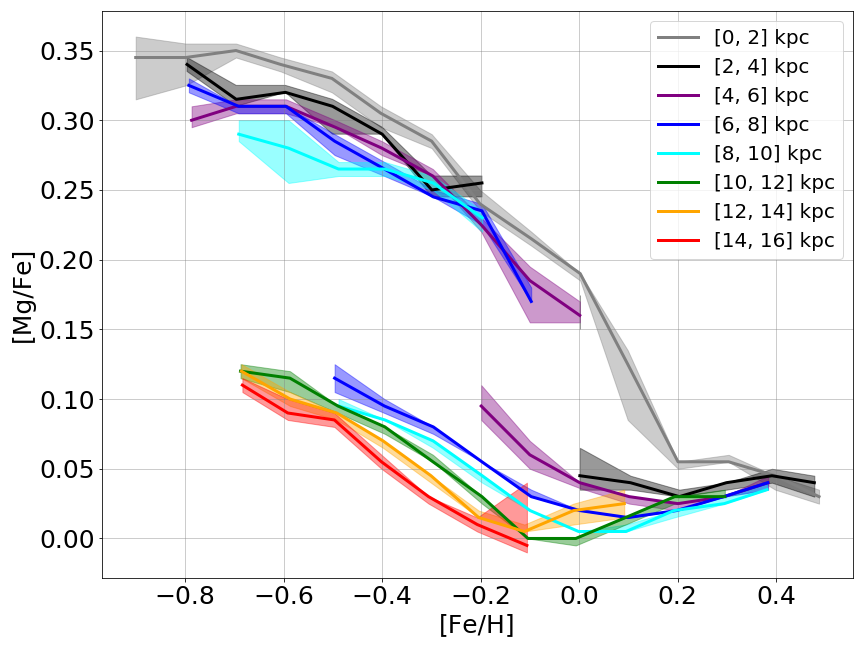}
   \includegraphics[width=\hsize]{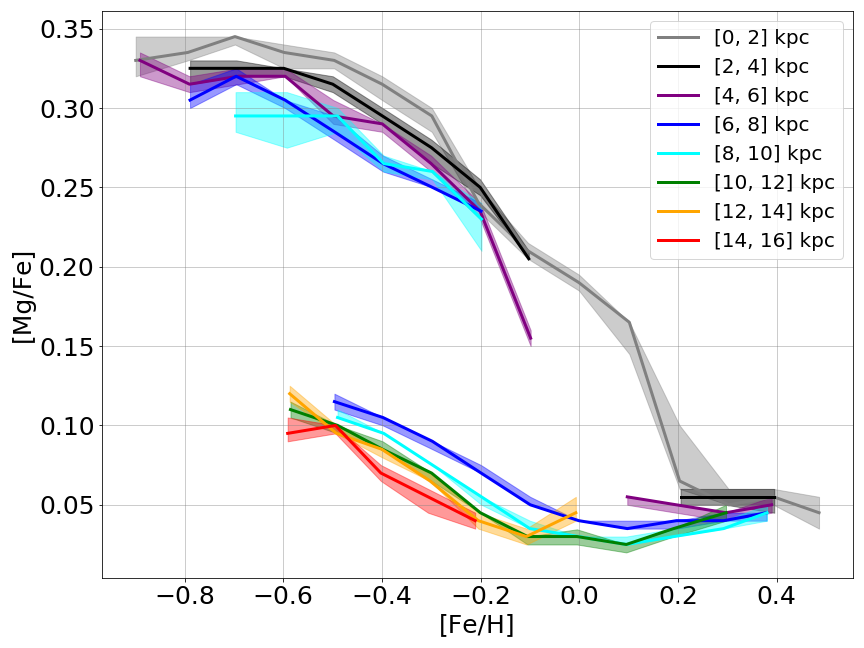}
   \includegraphics[width=\hsize]{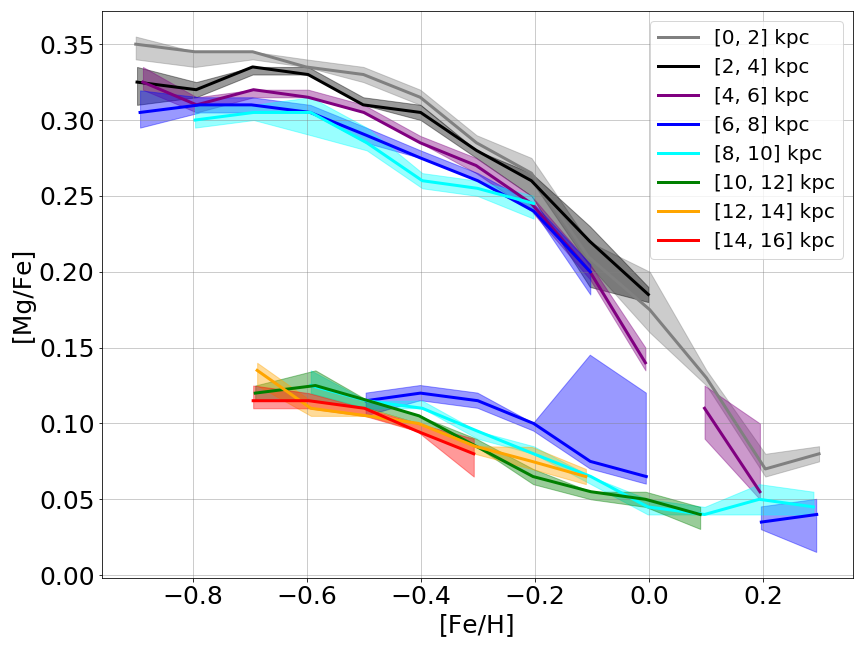}
   \caption{Same as Fig.~\ref{fig:fehRidge}, for different intervals of distance to the Galactic plane: $|Z| \in [0, 500]$~pc (top), $|Z| \in [500, 1000]$~pc (middle) and $|Z| \in [1000, 2000]$~pc (bottom).}
   \label{fig:appE2}
\end{figure}

For a given line of sight\footnote{outside of the Galactic plane}, looking at a larger distance from the Sun also means looking further out of the Galactic plane. Our main sample is made of stars selected in a broad $\pm 2$~kpc layer around the Galactic plane. In order to check that the $\FeH$-$\MgFe$ ridge line patterns were not resulting from a correlation in the selection function between the guiding radius $R_g$ and the Galactocentric vertical coordinate $Z$, we repeated the analysis conducted in Sect.~\ref{sect:FeHMgFe}, using narrower intervals of distance to the Galactic plane. Fig.~\ref{fig:appE1} and \ref{fig:appE2} reproduce respectively Fig.~\ref{fig:fehMosaic} and \ref{fig:fehRidge}, for $|Z| \in [0, 500]$~pc (top), $|Z| \in [500, 1000]$~pc (middle) and $|Z| \in [1000, 2000]$~pc (bottom).

In the three different vertical intervals, the pattern of the thick disc ridge lines is similar to the one shown in fig.~\ref{fig:fehRidge} for the main sample considered as a whole: i.e. the ridge lines are closely grouped and present a drift in $\MgFe$. In the closest interval to the plane ($|Z| < 0.5$~kpc), the thin disc behaviour is also globally similar to the one observed in Sect.~\ref{sect:FeHMgFe}: i.e. at sub-solar metallicity, the thin disc ridge lines follow parallel paths wich converge and overlap at super-solar metallicity. However, in the top panel of Fig.~\ref{fig:appE1}, in the guiding radius interval $R_g \in [4, 6]$~kpc, the thin disc ridge line extends to lower metallicity than in Fig.~\ref{fig:fehMosaic} and is disjoint from the thick disc ridge line. It is most likely the smaller contribution of the thick disc close to the plane which increases the contrast and visibility of the low-alpha sub-solar metallicity stars. At greater distance from the plane, the proportion of thin disc stars decreases and the shift between the ridge lines at sub-solar metallicity becomes less visible, especially in the farthest interval ($1 < |Z| < 2$~kpc).

%
%
\section{Chemical patterns in the inner 2~kpc\label{app:F}}
\begin{figure}[h!]
   \centering
   \includegraphics[width=\hsize]{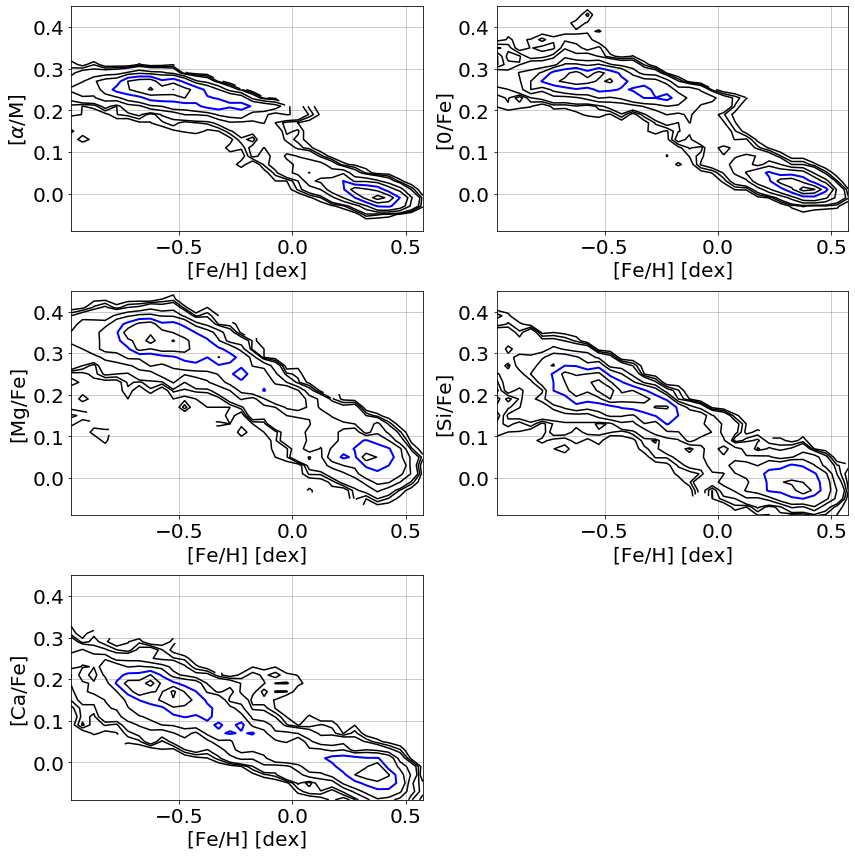}
   \caption{$\AM$, $\OFe$, $\MgFe$, $\SiFe$ and $\CaFe$ versus $\FeH$ distributions of the stars in the guiding radius interval $R_g \in [0, 2]$~kpc. The iso-density contours are plotted at 5, 7.5, 10, 20, 30, 70, 90 (black lines) and 50\% (blue line) of the maximum density.}
   \label{fig:appF}
\end{figure}

There is not yet unanimity on the morphology of the ($\FeH$, $\AFe$) distribution in the inner regions of the galaxy. Some studies report one sequence \citep[][as well as this study]{Bensby2017, Zasowski2019, Bovy2019, Lian2020c, Lian2020d} and others observe two \citep{RojasArriagada2019, Queiroz2020}. The comparison is not always straightforward, as these results could be based on spectra collected with different instruments and/or on different alpha-elements measured with different techniques. In this appendix we compare different APOGEE DR16 abundance ratios. Figure~\ref{fig:appF} presents the metallicity/alpha-ratio distributions of the stars contained in the innermost guiding radius bin, i.e. $R_g \in [0, 2]$~kpc. Five abundance ratios are shown: $\AM$, $\OFe$, $\MgFe$, $\SiFe$ and $\CaFe$. The two first, $\AM$ and $\OFe$, presents two sequences (connected by what could be the metal-poor tail of the low-alpha sequence). The three others, $\MgFe$, $\SiFe$ and $\CaFe$ show a single sequence. The processing and properties of the APOGEE SDSS DR16 data are described in \citet{Jonsson2020}. The five abundance ratios discussed in this appendix were derived by the ASPCAP pipeline, but not exactly with the same method. The {\it abundance parameter} $\AM$ was derived by global fitting of the full spectrum, while the individual abundances of ${\rm O}$, ${\rm Mg}$, ${\rm Si}$, ${\rm Ca}$ and ${\rm Fe}$ were measured using selected spectral windows and are expected to be more precise. In Fig.~\ref{fig:appF}, the double sequence shape of the $\AM$ and $\OFe$ versus $\FeH$ distributions is enhanced by the {\it finger} feature around $\FeH \sim 0.0$~dex and $\AM$/$\OFe \sim +0.2$~dex. This feature is also visible in the full APOGEE catalogue and is discussed in \citet{Jonsson2020}. It seems to be made of a small fraction of the cool ($\Teff < 4000$~K) super-solar metallicity giants. The nature of this sub-group, processing artefact or real physical feature, is not yet known, but \citet{Jonsson2020} recommend to be cautious when using these stars.

The fact that different alpha-elements behave differently, raises the question of which one to use. In this study, we decided to use magnesium because it is described in \citet{Jonsson2020} as the most precise abundance (together with silicon) in DR16.

%
%
\section{($\FeH$, $\MgFe$) ridges versus radius\label{app:G}}
\begin{figure}[h!]
   \centering
   \includegraphics[width=\hsize]{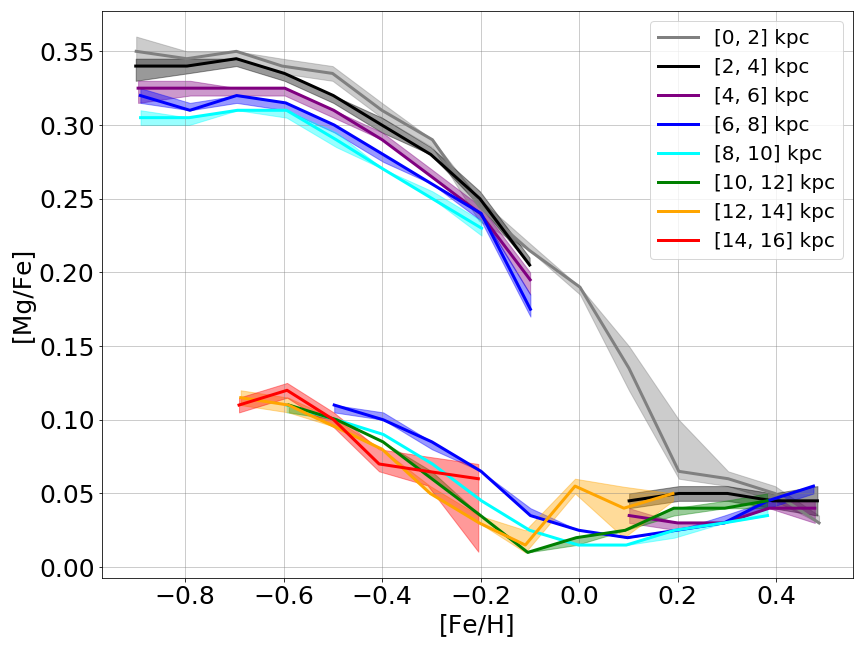}
   \caption{Ridge lines of the ($\FeH$, $\MgFe$) sequences for different Galactic radius annulus, from $[0, 2]$~kpc (gray) to $[14, 16]$~kpc (red). The shaded areas delimit the $\pm 1\sigma$ uncertainties (derived by bootstrap) on the estimates of the modes.}
   \label{fig:appG2}
\end{figure}

Figure~\ref{fig:appG2} shows the ridge lines of the ($\FeH$, $\MgFe$) sequences for different intervals of Galactic radius. 
Overall, it appears quite similar to Fig.~\ref{fig:fehRidge}, which displays the ridge lines as a function of guiding radius. Here too, the thick disc sequences are slightly shifted to lower $\MgFe$, proportionally to their radius. In the thin disc, at super-solar metallicity, the ridge lines defined by Galactic radius intervals are more scattered than those defined by guiding radius intervals. At sub-solar metallicity, the separations between the $[6, 8]$, $[8, 10]$ and $[10, 12]$~kpc sequences are still well visible, but less so at larger Galactic radius, where the dynamical blurring likely {\it hides} the shifts between the sequences.\\


%
%
\section{(Age, $\MgFe$) ridges versus radius\label{app:H}}
\begin{figure}[h!]
   \centering
   \includegraphics[width=\hsize]{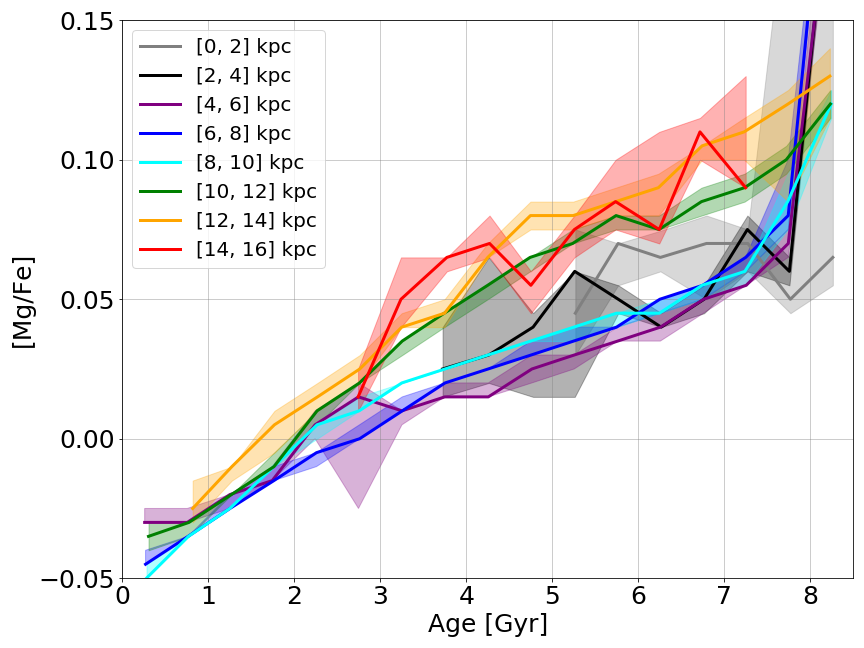}
   \caption{Ridge lines of the (Age, $\MgFe$) sequences for different Galactic radius annulus, from $[0, 2]$~kpc (gray) to $[14, 16]$~kpc (red). The shaded areas delimit the $\pm 1\sigma$ uncertainties (derived by bootstrap) on the estimates of the modes.}
   \label{fig:appH2}
\end{figure}

Figure~\ref{fig:appH2} shows the relative locations of the ridge lines of the age-$\MgFe$ thin disc sequences, for different interval in Galactic radius, from $R \in [0, 2]$~kpc (gray) to $[14, 16]$~kpc (red). The ridge lines are noisier than their counterpart calculated per interval of guiding radius (see Sect.~\ref{sect:ageMgFe}), but the main features seen in Fig.~\ref{fig:ageRidge} are still present. The ridge lines beyond $R = 10$~kpc present a steeper gradient than the ridge lines below. The difference of slope produces a gap between the two groups, beyond $\sim 2-3$~Gyr.\\


\end{appendix}
\end{document}